\documentclass[11pt,draftcls,onecolumn]{IEEEtran}
\usepackage{graphicx}
\usepackage{bm}
\usepackage{url}
\usepackage{mystyle}
\usepackage{amsfonts,amssymb,amsmath}
\allowdisplaybreaks
\def\defeq{\stackrel{\triangle}{=}}
\def\bI{{\mathbf I}}                
\def\bH{{\mathbf H}}                
\def\bP{{\mathbf T}}                
\def\bx{{\mathbf x}}                
\def\by{{\mathbf y}}                
\def\bn{{\mathbf n}}                
\def\bW{{\mathbf W}}                
\def\prob{{\mathbb P}}              
\def\expect{{\mathbb E}}            
\def\balpha{{\boldsymbol{\alpha}}}  

\title{Diversity of MIMO Linear Precoding}

\author{Ahmed Hesham Mehana and Aria Nosratinia\thanks{The authors are with the Department of
Electrical Engineering, The University of Texas at Dallas,
Richardson, TX 75083-0688 USA, E-mail: ahmed.mehana@student.utdallas.edu;
aria@utdallas.edu.}}

\begin{document}

\maketitle

\begin{abstract}
Linear precoding is a relatively simple method of MIMO signaling that
can also be optimal in certain special cases. This paper is dedicated
to high-SNR analysis of MIMO linear precoding. The
Diversity-Multiplexing Tradeoff (DMT) of a number of linear precoders
is analyzed. Furthermore, since the diversity at finite rate (also
known as the fixed-rate regime, corresponding to multiplexing gain of
zero) does not always follow from the DMT, linear precoders are also
analyzed for their diversity at fixed rates. In several cases, the
diversity at multiplexing gain of zero is found not to be unique, but
rather to depend on spectral efficiency. The analysis includes the
zero-forcing (ZF), regularized ZF, matched filtering and Wiener
filtering precoders. We calculate the DMT of ZF precoding under two
common design approaches, namely maximizing the throughput and
minimizing the transmit power. It is shown that regularized ZF (RZF)
or Matched filter (MF) suffer from error floors for all positive
multiplexing gains.  However, in the fixed rate regime, RZF and MF
precoding achieve full diversity up to a certain spectral efficiency
and zero diversity at rates above it. When the regularization
parameter in the RZF is optimized in the MMSE sense, the structure is
known as the Wiener precoder which in the fixed-rate regime is shown
to have diversity that depends not only on the number of antennas, but
also on the spectral efficiency.  The diversity in the presence of
{\em both} precoding and equalization is also analyzed.
\end{abstract}

\section{Introduction}

Precoding is a preprocessing technique that exploits channel-state
information at the transmitter (CSIT) to match the transmission to the
instantaneous channel
conditions~\cite{ProakisShamai:IT98,Scaglione:JSP02,Jayaweera:TIF03}.
Linear and non-linear precoding designs are available in the
literature~\cite{Joham:JSP05}.  Linear precoding in particular
provides a simple and efficient method to utilize CSIT. Linear
precoding has been shown to be optimal in certain situations involving
partial CSIT~\cite{Caire:TIF99, Skoglund:SAC03}, however, in many
instances the main motivation of linear precoders is to simplify the MIMO
receiver.

Linear precoders include zero-forcing (ZF), matched filtering (MF),
Wiener filtering, and regularized zero-forcing (RZF). The ZF precoding
schemes were extensively studied in multiuser systems as the ZF
decouples the multiuser channel into independent single-user channels
and has been shown to achieve a large portion of dirty paper coding
capacity~\cite{Taesang:SAC06}.  ZF precoding often involves {\em
  channel inversion}, using the pseudo-inverse of the channel or other
generalized inverses~\cite{Joham:JSP05}.  
Matched filter (MF)
precoding~\cite{Esmailzadeh:ICC93}, similarly to the MF receiver, is
interference limited at high SNR but it outperforms the ZF precoder at
low SNR~\cite{Joham:JSP05}.  The regularized ZF precoder, as the name
implies, introduces a regularization parameter in channel inversion. If
the regularization parameter is inversely proportional to SNR, the RZF
of~\cite{Hochwald:TCOMM05} is identical to the Wiener
filter precoding~\cite{Joham:JSP05}.
Peel et al.~\cite{Hochwald:TCOMM05} introduce a vector perturbation
technique to reduce the transmit power of the RZF method, showing that
in this way RZF can operate near channel capacity.

This paper analyzes the diversity of MIMO linear precoding with or
without linear receivers. 
We show that a MIMO ZF precoder with a maximum likelihood receiver
has minimal spatial diversity, and that Wiener precoders produce a
diversity that is a complex function of spectral efficiency and the
number of transmit and receive antennas. 
At very low rates, the Wiener precoder enjoys a maximal diversity which
is the product of the number of transmit and receive antennas, while
at very high rates it achieves a minimal diversity which is the same
as ZF diversity. These results are reminiscent of
MIMO linear equalizers~\cite{Hesham:ISIT10}, even though in general
the behavior of equalizers (receive side) can be very different from
precoders (transmit side) and the analysis does not carry
from one to the other.
We also show that MIMO systems with RZF and MF precoders (together
with optimal receivers) exhibit a new kind of rate-dependent diversity
that has not to date been observed or reported, i.e., they either have
full diversity or zero diversity (error floor) depending on the
operating spectral efficiency $R$.

We also provide DMT analysis for all precoders mentioned above. 
The fact that DMT and the diversity under fixed-rate regime require
separate analyses has been established for MIMO linear
equalizers~\cite{Hedayat:JSP07,Hesham:ISIT10} and is by now a
well-understood phenomenon. Essentially, the reason is that various
fixed rates (spectral efficiencies) for MIMO precoding result
in distinctly different diversities, whereas DMT analysis assigns
only a single value of diversity to all fixed rates (all fixed rates
correspond to multiplexing gain zero).

\begin{remark}
Due to symbolic similarities, it may be tempting to draw the
conclusion that if $d(r)$ is the diversity at multiplexing gain $r$,
then substituting $r=0$ in the same mathematical expression will give
the diversity at multiplexing gain zero $d(0)$. However, despite
appearances, there is no solid relationship between $d(r)$ and
$d(0)$. The standard DMT arguments are based on the seminal work of
Zheng and Tse~\cite{Zheng:JIT03} whose developments depend critically
on the positivity of $r$. For example, the proof
of~\cite[Lemma~5]{Zheng:JIT03} depends critically on $r$ being
strictly positive.
More importantly, the
asymptotic outage calculations in~\cite[p.~1079]{Zheng:JIT03}
implicitly use $r>0$ and result in the outage region:
\[
{\mathcal A} = \{\alpha: \sum_i (1-\alpha_i)^+ <r \}
\]
where $\alpha_i$ are the exponential order of the channel eigenvalues,
i.e., $\lambda_i=\rho^{-\alpha_i}$. If we set $r=0$ this expression
implies that the outage region is always empty, which is clearly not true.

Thus, the DMT as calculated by the standard methods
of~\cite{Zheng:JIT03} does not extend to $r=0$. The DMT $d(r)$ is {\em
  sometimes} continuous at zero, including e.g. the examples
in~\cite{Zheng:JIT03}, but continuity at $r=0$ does not always
hold. In fact, there are systems where $d(0)$, the diversity at
multiplexing gain zero, is not even uniquely defined. It is possible
for diversity to take multiple values as a function of rate $R$. This
fact has been observed and analyzed, e.g.,
in~\cite{Hedayat:JSP07,Hesham:ISIT10,Kumar:JIT09}.  The work in the
present paper also produces several examples of this phenomenon.
\end{remark}

This paper is organized as follows. Section~\ref{sec:SysMod} describes
the system model. Section~\ref{sec:PrecOnly} provides outage analysis
of many precoded MIMO systems.  Section~\ref{sec:DMTPrecOnly} provides
the DMT analysis.  The case of joint linear transmit and receive
filters is discussed in Section~\ref{sec:JointLinearTxRx}.
Section~\ref{sec:SimRes} provides simulations that illuminate our
results.

 \begin{figure*}
\centering
\includegraphics{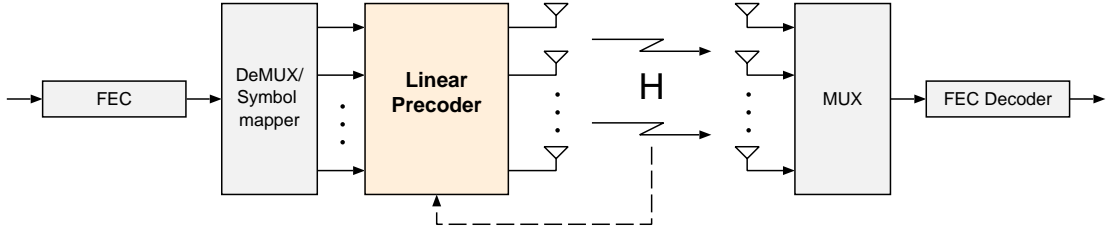}
\caption{MIMO with linear precoder}
\label{fig:LinearPrecoder}
\end{figure*}

\section{System Model}
\label{sec:SysMod}

A MIMO system with linear precoding is depicted in
Fig.~\ref{fig:LinearPrecoder}. This system uses the linear precoder to
manage the interference between the streams in a MIMO system to avoid
a lattice decoder in the receiver. We consider a flat fading channel
$\bH \in \mathbb{C}^{N\times M}$, where $M$ and $N$ are the number of
transmit and receive antennas, respectively. While $M\geqslant N$ when
using linear precoding alone, we have $N\geqslant M$ or $M\geqslant N$
when using precoding together with receive-side linear equalization
depending on whether the precoder is designed for the equalized
channel or the equalizer is designed for the precoded channel (see
Figure~\ref{fig:linearprecoderEq}). The input-output system model for
flat fading MIMO channel with $M$ transmit and $N$ receive antennas is
given by
\begin{equation}
 \by = \bW\bH \bP\bx + \bW\bn \label{eq:SysMod}
\end{equation}
where $\bP \in \mathbb{C}^{M\times B}$ is the precoder matrix, $\bW
\in\mathbb{C}^{B\times N}$ is the receiver side equalizer. The latter
may be set to identity in cases where the receiver does not use linear
equalization. The number of information symbols is $B\leqslant \min
(M,N)$, the transmitted vector is $ {\bx \in
  \mathbb{C}^{\ B \times 1}}$, and ${\bn \in \mathbb{C}^{\ N \times
    1}} $ is the Gaussian noise vector. The vectors ${\bx}$ and
${\bn}$ are assumed independent.

We aim to characterize the diversity gain, $ d(R,M,N)$, as a function
of the spectral efficiency $R$ (bits/sec/Hz) and the number of
transmit and receive antennas. This requires a Pairwise Error
Probability (PEP) analysis which is not directly tractable. Instead,
we find the exponential order of outage probability and then
demonstrate that outage and PEP exhibit identical exponential orders.

The objective of linear precoding (possibly together with linear
equalization at the receiver) is to transform the MIMO channel into
$\min(M,N)$ parallel channels that can be described by
\begin{equation}
y_{k}=\sqrt{\gamma_{k}}x_{k}+n_{k}, \quad k=1,\dots,B \label{eq:ParallelCh}
\end{equation}
where $\gamma_{k}$ is the SINR at the $k$-th receiver output and
$B=min(M,N)$.  
Following the notation of~\cite{Tajer:WCOM10}, we define the
outage-type quantities
\begin{align}
 P_{out}(R,N,M) &\triangleq  \prob(I(\bx;\by)<R)
\label {eq:OutageProb}\\
 d_{out}(R,N,M) &\triangleq  -\lim_{\rho\rightarrow\infty}\frac{\log P_{out}(R,M,N)}{\log \rho}
\end{align}
where $\rho$ is the transmitted equivalent SNR.


The outage probabilities of MIMO systems under joint spatial encoding
is respectively given by~\cite{Hedayat:JSP07,Kumar:JIT09}
\begin{align}
&P_{\text{out}}\triangleq \prob\bigg( \sum_{k=1}^{B}\log (1+\gamma_{k}) \leqslant R\bigg)\label{eq:JointEncOut}
\end{align}

We shall perform outage analysis for different precoders/equalizers as
the first step towards deriving the diversity function. We then
provide lower and upper bounds on error probability via outage
probabilities. This two-step approach was first proposed
in~\cite{Zheng:JIT03} due to the intractability of the direct PEP
analysis for many MIMO architectures.

We denote the {\em exponential equality} of two functions $f(\rho)$
and $g(\rho)$ as $f(p) \doteq g(p)$ when
\begin{equation}
\lim_{\rho \rightarrow \infty}
\frac{\log f(\rho)}{\log(\rho)}=\lim_{\rho \rightarrow \infty}
\frac{\log g(\rho)}{\log(\rho)} \nonumber
\end{equation}

In the following, we shall need to specify various upper and lower
bounds or approximations of the SINR $\gamma$, which
will give rise to a number of pseudo-SINR variables $\hat{\gamma}$,
$\breve{\gamma}$, and $\bar{\gamma}$.

\section{Precoding Diversity}
\label{sec:PrecOnly}
In this section we analyze a linearly precoded MIMO system where $M
\ge N$ and the number of data streams $B$ is equal to $N$.

\subsection{Zero-Forcing Precoding}
\label{sec:ZFPrecGen}

The ZF precoder completely eliminates the interference at the
receiver. ZF precoding is well studied in the literature via
performance measures such as throughput and fairness under a total (or
per antenna) power constraint~\cite[and references
  therein]{Wiesel:JSP08}.

\subsubsection{Design Method I}
\label{sec:ZFPrec}

 One approach to design the ZF precoder
is to solve the following problem~\cite{Joham:JSP05}
\begin{align}
\bP= &\arg\min_{\bP} \expect \big [ ||\bP\bx||^{2}_{2}  \big ] \label{eq:OptimizeZFPower}\\& \text{subject to  } \bH\bP=\bI  \nonumber
\end{align}
The resulting ZF transmit filter is given by
\begin{equation}
\bP = \beta \,\bH^H(\bH\bH^H)^{-1} \in \mathbb{C}^{M\times N} \label{eq:ZFFirstDesignTotPower}
\end{equation}
where $\beta$ is a scaling factor to satisfy the transmit power constraint, that is~\cite{Joham:JSP05}
\begin{equation}
\beta^2\text{tr} \big( \bP \bP^H  \big)\leqslant \rho \label{eq:BetaRhoZF}
\end{equation}
where we assume that the noise power is one and that the information streams are independent.
From~\eqref{eq:BetaRhoZF}, the received SINR per stream is thus given by
\begin{equation}
\gamma^{ZFP}_{k}=\frac{\rho}{\text{tr}(\bH\bH^H)^{-1}}.\label{eq:SINRZF}
\end{equation} 
Using~\eqref{eq:JointEncOut}, the outage probability is given by
\begin{equation}
P_{\text{out}}= \prob\bigg( N\log \big(1+\frac{\rho}{\text{tr}((\bH\bH^H)^{-1})}\big) \leqslant R\bigg)\label{eq:JointEncOutZF1}
\end{equation}

A direct evaluation of~\eqref{eq:JointEncOutZF1} is intractable since
the diagonal elements of $(\bH\bH^H)^{-1}$ are distributed according to
the inverse-chi-square
distribution~\cite{Heath:ISIT02,Hedayat:JSP07}. We instead
bound~\eqref{eq:JointEncOutZF1} from below and above and show that the
two bounds match asymptotically.


Let $\lambda_k$ be the k-th eigenvalue of $\bH\bH^H$. Equation~\eqref{eq:JointEncOutZF1} can be written as
\begin{equation}
P_{\text{out}}= \prob\bigg( N\log \big(1+\frac{\rho}{\sum_{k=1}^{N}\frac{1}{\lambda_k}}\big) \leqslant R\bigg)\nonumber
\end{equation}
which can be bounded as
\begin{align}
P_{\text{out}}&\leqslant \prob\bigg( N\log (1+\frac{\rho}{N}\lambda_{\text{min}}) \leqslant R\bigg)\label{eq:OutProbUBZFFirstDes}\\
&= \prob\bigg( \lambda_{\text{min}} \leqslant N (2^{\frac{R}{N}}-1) R \rho^{-1}\bigg)\nonumber\\
&\dot{=} \;\,\prob\big( \lambda_{\text{min}} \leqslant \rho^{-1}\big)\label{eq:ZFPrecOutLBLambdaMin}.
\end{align}

The marginal distribution $f_1(\lambda)$ of $\lambda_{\text{min}}$ is
$f_1(\lambda) = c \lambda^{(M-N)}$~\cite{Telatar:EUR99} where $c$ is a
constant, therefore the bound in~\eqref{eq:ZFPrecOutLBLambdaMin} can
be evaluated~\cite{Kumar:JIT09} yielding:
\begin{equation}
P_{\text{out}}\;\dot{\leqslant}\;\; \rho^{-(M-N+1)}\label{eq:JointEncZFUBResult}.
\end{equation}

We now proceed with a lower bound on outage.  The outage probability
in~\eqref{eq:JointEncOutZF1} can be bounded:
\begin{align}
P_{\text{out}}&= \prob\bigg( N\log (1+\frac{\rho}{\text{tr}(\bH\bH^H)^{-1}}) \leqslant R\bigg)\nonumber\\
&\geqslant \prob\bigg( N\log (1+\frac{\rho}{(\bH\bH^H)^{-1}_{kk}}) \leqslant R\bigg)\nonumber\\
&\dot{=}\;\; \prob\bigg(  z \leqslant \rho^{-1}\bigg)\label{eq:JointEncOutZFLB2}
\end{align}
where we have made a change of variable $z=\frac{1}{(\bH\bH^H)^{-1}_{kk}}$. 

 The random variable $z$ in~\eqref{eq:JointEncOutZFLB2} is distributed
 according to the chi-square distribution with $2(M-N+1)$ degree of
 freedom, i.e. $z\sim\mathcal{X}^22(M-N+1)$~\cite{Heath:ISIT02}.  Thus
 the bound in~\eqref{eq:JointEncOutZFLB2} can be evaluated~\cite{Hedayat:JSP07} yielding:
\begin{equation}
P_{\text{out}}\;\dot{\geqslant}\;\; \rho^{-(M-N+1)}\label{eq:JointEncZFLBResult}.
\end{equation}

From~\eqref{eq:JointEncZFUBResult} and~\eqref{eq:JointEncZFLBResult},
we conclude that the diversity of MIMO system using the ZF precoder
given by~\eqref{eq:OptimizeZFPower} and joint spatial encoding is
\begin{equation}
d^{ZFP}=M-N+1.\label{eq:DivZFPrecJE}
\end{equation}

\subsection{Zero-Forcing Precoding: Design Method II}
\label{sec:ZFPrec2}

Notice that the ZF precoder design in~\eqref{eq:OptimizeZFPower}
minimizes the transmitted power. Another approach for ZF precoding
design allocates unequal power levels across the transmit antennas to
optimize some performance measure. For instance, consider the
optimization problem~\cite{Wiesel:JSP08}
\begin{align}
&\max_{p_k,\bP} \;\;f(p_k)\nonumber\\
&\text{subject to} \;\;\;\;\; \bH\bP=\text{diag}\{\sqrt{p_1},\dots,\sqrt{p_{M}}\}\nonumber\\
 &\hspace{26pt} \expect||\bP\bx||^2\leqslant \rho \label{eq:ZFGeneral}
\end{align}
where $f(p_k)$ is an arbitrary function of the transmitted power $p_k$
on the $k$-th antenna.

The optimal solution for~\eqref{eq:ZFGeneral} (assuming independent
transmit signaling) is given by~\cite[Theorem 1]{Wiesel:JSP08}:
\begin{equation}
 \bP = \bH^H(\bH\bH^H)^{-1}\text{diag}\{\sqrt{p_1},\dots,\sqrt{p_{M}}\}
\label{eq:ZFPrecThroOptimal}
\end{equation}
 where $p_k$ is obtained by solving
\begin{align}
&\max_{p_k} \;\;f(p_k)\nonumber\\
&\text{subject to} \;\;\;\;\; \sum_{k}p_k\big[ (\bH\bH^H)^{-1} \big]_{kk}\leqslant \rho \label{eq:ZFPowerGeneral}
\end{align}
In our case, we maximize the throughput, therefore
$f(p_k)=\sum_{k}\log (1+\gamma^{ZFP}_{k})$. After setting the
derivatives of the appropriate Lagrangian function to zero, the
solution of the power allocation problem in~\eqref{eq:ZFPowerGeneral}
is given by
\begin{equation}
p_{k}=\frac{\rho+\sum_{k=1}^{M}(\bH\bH^H)^{-1}_{kk}}{M(\bH\bH^H)^{-1}_{kk}}-1\label{eq:SINRZFThroOptimal}
\end{equation}


Substituting~\eqref{eq:SINRZFThroOptimal}
in~\eqref{eq:JointEncOut}, the outage probability is given by
\begin{align}
P_{\text{out}}&= \prob\bigg( \sum_{k=1}^{N}\log (\frac{\rho+\sum_{k=1}^{N}(\bH\bH^H)^{-1}_{kk}}{M(\bH\bH^H)^{-1}_{kk}}) \leqslant R\bigg)\nonumber\\
&\dot{=}\;\;\prob\bigg( \sum_{k=1}^{N}\log (\frac{\rho}{M(\bH\bH^H)^{-1}_{kk}}) \leqslant R\bigg)\label{eq:ZFThroOptimal1}\\
&= \prob\bigg( \sum_{k=1}^{N}\log (\frac{M(\bH\bH^H)^{-1}_{kk}}{\rho}) \geqslant -R\bigg)\\
&\leqslant \prob\bigg(N\log \sum_{k=1}^{N} (\frac{(\bH\bH^H)^{-1}_{kk}}{\rho}) \geqslant -R\bigg)\label{eq:ZFThroOptimal2}\\
&= \prob\bigg( \sum_{k=1}^{N} \frac{1}{\rho\lambda_k} \geqslant  2^{-\frac{R}{N}}\bigg)\nonumber\\
&\leqslant \prob\bigg(  \frac{1}{\rho\lambda_{\text{min}}} \geqslant \frac{1}{N}  2^{-\frac{R}{N}}\bigg) \;  \nonumber\\
&\doteq \; \prob\bigg(  \lambda_{\text{min}} \leqslant  \rho^{-1}\bigg)    \label{eq:ZFThroOptimal3}\\
&\doteq \rho^{-(M-N+1)}  \label{eq:ZFThroOptimal4}.  
\end{align}
where the exponential equality~\eqref{eq:ZFThroOptimal1} holds at high
SNR,~\eqref{eq:ZFThroOptimal2} follows from Jensen's inequality, and
the transition from~\eqref{eq:ZFThroOptimal3}
to~\eqref{eq:ZFThroOptimal4} again due to the marginal distribution of
$\lambda_{\min}$ via the method of~\cite{Kumar:JIT09}.

A lower bound on the outage probability can be given as follows.
Starting with~\eqref{eq:ZFThroOptimal1} and using Jensen's inequality
we have
\begin{align}
P_{\text{out}}&\dot{=}\;\;\prob\bigg( \sum_{k=1}^{N}\log (\frac{\rho}{N(\bH\bH^H)^{-1}_{kk}}) \leqslant R\bigg)\nonumber\\
&\geqslant \prob\bigg(N\log \frac{1}{N^2} \sum_{k=1}^{N} \frac{\rho}{(\bH\bH^H)^{-1}_{kk}} \leqslant R\bigg).\label{eq:ZFThroOptimaLB1}
\end{align}

The singular value decomposition of $\bH$ and the corresponding eigen
decomposition of $\bH\bH^H$ are given by
\begin{align}
&\bH\quad\hspace{6pt}=\bU\,\Gamma\,\bV^H\nonumber\\ 
&\bH\bH^H=\bU\Lambda\,\bU^H \nonumber
\end{align}
 where $\bU\in\mathbb{C}^{N\times N}$ and
$\bV \in\mathbb{C}^{M\times M}$ are unitary matrices,
$\Gamma\in\mathbb{R}^{N\times M}$ is a rectangular matrix with
non-negative real diagonal elements and zero off-diagonal elements,
and $\Lambda=\Gamma\Gamma^T\in\mathbb{R}^{N\times N}$ is a diagonal
matrix whose diagonal elements are the eigenvalues of $\bH\bH^H$. Let
$\bu_k$ be the $k$-th column of $\bU^H$. We have
\begin{equation}
(\bH\bH^H)^{-1}_{kk}=\bu^H_k\Lambda^{-1}\bu_k
=\sum_{l=1}^{N}\frac{|u_{kl}|^2}{\lambda_l}\label{eq:GetInvEigDecomp}
\end{equation}
where $u_{kl}$ is the $(k,l)$ entry of the matrix $\bU$.

The bound in~\eqref{eq:ZFThroOptimaLB1} can be rewritten  
\begin{align}
P_{\text{out}} \; &\dot{\geqslant} \;  \prob\bigg(N\log \frac{1}{N^2} \sum_{k=1}^{N} \frac{1}{\sum_{l=1}^{N}\frac{|u_{kl}|^2}{\rho\lambda_l}} \leqslant R\bigg)\nonumber\\
 &\geqslant   \prob\bigg(N\log \frac{1}{N^2} \sum_{k=1}^{N} \frac{1}{\sum_{l=1}^{N}\frac{|u_{kl}|^2}{1+\rho\lambda_l}} \leqslant R\bigg)\label{eq:ZFThroOptimalLB2}
\end{align}

We can lower bound the probability in~\eqref{eq:ZFThroOptimalLB2} in
by observing that the term
\[
\frac{1}{N}\sum_{k=1}^{N}
\frac{1}{\sum_{l=1}^{N}\frac{|u_{kl}|^2}{1+\rho\lambda_l}}
\] 
is similar
to~\cite[Eq.(18)]{Kumar:JIT09}, thus the analysis of~\cite{Kumar:JIT09}
applies and we obtain
\begin{equation}
P_{\text{out}} \;\dot{\geqslant}\;  \prob\bigg(  \lambda_{\text{min}} \leqslant  \rho^{-1}\bigg) = \rho^{-(M-N+1)} \label{eq:OutProbZFSecondDesign}.
\end{equation}

Thus, the MIMO ZF precoding with unequal power
allocation~\eqref{eq:ZFPowerGeneral} achieves diversity order $M-N+1$.

Recall that the diversity is defined based on the error
probability. In Appendix~\ref{appendix:PEP} we provide the pairwise
error probability (PEP) analysis for the zero-forcing and regularized
zero-forcing precoded systems and show that the outage and error
probabilities exhibit same diversity.


\subsection{Regularized Zero-Forcing Precoding}
\label{sec:RZFPrec}

In general, direct channel inversion performs poorly due to the
singular value spread of the channel
matrix~\cite{Hochwald:TCOMM05}. One technique often used is to
regularize the channel inversion:
\begin{equation}
\bP=\beta\,\bH^H(\bH\bH^H+c\,\bI)^{-1} \label{eq:RZFPrecoderFilter}
\end{equation}
where $\beta$ is a normalization factor and $c$ is a fixed constant.

Recall 
\begin{align}
\by = \bH\bP\bx+\bn&=\beta\bU\Lambda(\Lambda+c\,\bI)^{-1}\bU^H\bx+\bn \label{eq:EigDecSysMod}
\end{align}
allowing us to decompose the received waveform at each antenna 
 into signal, interference, and noise terms:
\begin{align}
y_k &=\beta\bigg( \sum_{l=1}^{N} \frac{\lambda_l}{\lambda_l+c\,} |u_{kl}|^2 \bigg) x_k \;\;+\nonumber\\
 &\quad\; \beta\sum_{i=1,i\neq k}^{N} \bigg( \sum_{l=1}^{N} \frac{\lambda_l}{\lambda_l+c\,} u_{kl}u^*_{il} \bigg)  x_{i}  + n_{k}\label{eq:RecSig}
\end{align}
where the scaling factor $\beta$ is given by
$\beta=\frac{1}{\sqrt{\eta }}$ and
\begin{align}
 \eta&=\text{tr}\big[(\bH\bH^H+c\,\bI)^{-1}\bH\bH^{H}(\bH\bH^H+c\,\bI)^{-1}\big]\nonumber\\
&=\text{tr}\big[(\bU\Lambda\bU^H+c\,\bI)^{-1}\bU\Lambda\bU^H(\bU\Lambda\bU^H+c\,\bI)^{-1}\big]\nonumber\\
&=\text{tr}\big[\bU(\Lambda+c\,\bI)^{-1}\Lambda(\Lambda+c\,\bI)^{-1}\bU^H\big]\nonumber\\
&=\text{tr}\big[\Lambda(\Lambda+c\,\bI)^{-2}\big]=\sum_{l=1}^{N}\frac{\lambda_l}{(\lambda_l+c\,)^2}.\label{eq:NormFactor}
\end{align}
The received signal power is given by
\begin{align}
P_T&=E||\bH\bP\bx||^2\nonumber\\
&=E\bigg[ \beta^2\text{tr}\bigg( \bU\Lambda(\Lambda+c\,\bI)^{-1}\bU^H\bx \bx^H\bU(\Lambda+c\,\bI)^{-1}\Lambda \bU^H  \bigg) \bigg]\nonumber\\
&=E\bigg[ \beta^2\text{tr}\bigg(\Lambda(\Lambda+c\,\bI)^{-1}\bU^H\bx \bx^H\bU(\Lambda+c\,\bI)^{-1}\Lambda \bU^H\bU  \bigg) \bigg]\nonumber\\
&= \beta^2\text{tr}\bigg(\Lambda(\Lambda+c\,\bI)^{-1}\bU^H E(\bx \bx^H)\bU(\Lambda+c\,\bI)^{-1}\Lambda  \bigg) \nonumber\\
&= \frac{\beta^2 \rho}{N}\text{tr}\big[(\Lambda+c\,\bI)^{-2}\Lambda^2   \big] = \frac{\beta^2 \rho}{N}\sum_{l=1}^{N} \frac{\lambda_l^2}{(\lambda_l+c\,)^2}. \label{eq:TotSigPower}
\end{align}
where we have used $\expect(\bx \bx^H) = \frac{\rho}{N}\bI$.

The SINR is evaluated by computing the signal and interference powers
from~\eqref{eq:RecSig}. For a given channel $\bH$, the power of
desired and interference signals at the $k$-th receive antenna are
respectively given by
\begin{align}
P^{(k)}_D&=\frac{\beta^2\rho}{N}\bigg( \sum_{l=1}^{N} \frac{\lambda_l}{\lambda_l+c\,} |u_{kl}|^2 \bigg)^2\label{eq:SigPower}\\
P^{(k)}_I&=\frac{\beta^2\rho}{N}\sum_{i=1,i\neq k}^{N} \bigg| \sum_{l=1}^{N} \frac{\lambda_l}{\lambda_l+c\,} u_{kl} u^*_{il} \bigg|^2.   \label{eq:InterSigPower}
\end{align}

Thus the SINR for the $k$-th signal stream is given by
\begin{align}
\gamma_k&=\;\frac{P_D^{(k)}}{P_I^{(k)}+1}\nonumber\\
&=\frac{\frac{\beta^2\rho}{N}\bigg( \sum_{l=1}^{N} \frac{\lambda_l}{\lambda_l+c\,}|u_{kl}|^2  \bigg)^2} { \frac{\beta^2\rho}{N}\sum_{i=1,i\neq k}^{N} \bigg| \sum_{l=1}^{N} \frac{\lambda_l}{\lambda_l+c\,} u_{kl} u^*_{il} \bigg|^2 +1} \label{eq:SINRRZF}\\
\end{align}
recall $\eta$ is given by~\eqref{eq:NormFactor}.

Defining the exponential order of eigenvalues $\lambda_l=\rho^{-\alpha_l}$ in a manner similar to~\cite{Zheng:JIT03}, 
\begin{align}
&{\gamma}_{k}= \frac{\bigg( \sum_{l} \frac{\rho^{-\alpha_l}}{\rho^{-\alpha_l}+c\,} |u_{kl}|^2 \bigg)^2} { \sum_{i\neq k} \bigg| \sum_{l=1}^{N} \frac{\rho^{-\alpha_l}}{\rho^{-\alpha_l}+c\,} u_{kl} u^*_{il} \bigg|^2 +N\,\rho^{-1}\,\eta} \nonumber\\
&\dot{=}\; \frac{\bigg( \sum_{l} \rho^{-\alpha_l} |u_{kl}|^2 \bigg)^2} { \sum_{i\neq k} \bigg| \sum_{l=1}^{N} u_{kl} u^*_{il}\rho^{-\alpha_l}  \bigg|^2 +N\,\rho^{-1} \sum_{l=1}^{N}\rho^{-\alpha_l}}\label{eq:SINRRZFDef2}
\end{align}
where the asymptotic equality follows because in all terms $c$
dominates $\rho^{-\alpha_l}$, a fact that also implies $\eta \doteq \sum_l
\rho^{-\alpha_l}$.

Multiplying the numerator and denominator of~\eqref{eq:SINRRZFDef2} by
$\rho^{2}$, we have
\begin{align}
{\gamma}_{k}&\dot{=}\frac{\bigg( \sum_{l} \rho^{1-\alpha_l} |u_{kl}|^2 \bigg)^2} { \sum_{i\neq k} \bigg| \sum_{l=1}^{N} u_{kl} u_{il}^* \rho^{1-\alpha_l}  \bigg|^2 +\, N\sum_{l=1}^{N}\rho^{1-{\alpha}_l}}. \label{eq:SINRRZFDef3}
\end{align}
The sum in the
numerator of~\eqref{eq:SINRRZFDef3} is, in the SNR exponent, equivalent to:
\begin{align}
\sum_{l} \rho^{1-\alpha_l} |u_{kl}|^2 &\doteq \rho^{1-\alpha_{\min}} \sum_{l}  |u_{kl}|^2 \nonumber\\
&=\rho^{1-\alpha_{\min}} \label{eq:Aprox1}
\end{align}
where we use the fact that $ \sum_{l} |u_{kl}|^2=1$.
Similarly, for the first term in the denominator of~\eqref{eq:SINRRZFDef3}
\begin{align}
 \sum_{i\neq k} \bigg| \sum_{l=1}^{N} u_{kl} u_{il}^* \rho^{1-\alpha_l}  \bigg|^2 &\doteq \rho^{2-2\alpha_{\min}}  \sum_{i\neq k}  \bigg| \sum_{l=1}^{N} u_{kl} u_{il}^* \bigg|^2 \nonumber\\
&= \rho^{2-2\alpha_{\min}}  \sum_{i\neq k} w_{ki} \label{eq:Aprox2}
\end{align}
where we define $w_{ki} \triangleq  \bigg| \sum_{l=1}^{N} u_{kl} u_{il}^* \bigg|^2 $. Notice that $w_{ki} \le 1$.

Using~\eqref{eq:Aprox1} and~\eqref{eq:Aprox2}, the SINR
in~\eqref{eq:SINRRZFDef3} is given by
\begin{align}
\gamma_k &\dot{=} \frac{\bigg( \rho^{1-\alpha_{\min}} \bigg)^2} {
\rho^{2-2\alpha_{\min}}  \sum_{i\neq k} w_{ki} +\,
  N\sum_{l=1}^{N}\rho^{1-{\alpha}_l}}.\label{eq:SINRRZFDefC3}
\end{align}

If all $\alpha_\ell>1$ then the exponents of
$\rho$ are negative and the denominator is dominated by its second
term, which also dominates the numerator. If at least one of the
$\alpha_\ell\le 1$, then the maximum exponent which corresponds to
${\alpha}_{\min}$ dominates each summation. Thus we have:

\begin{equation}
{\gamma}_k  \doteq
 \begin{cases}
 \rho^{1-\alpha_{\min}} &\text{$\alpha_l>1\; \forall l$ } \\
 \frac{\big (\rho^{1-\alpha_{\min}}\big)^2  } {\rho^{2-2\alpha_{\min}}  \sum^N_{\substack{i=1 \\ i\neq k}} w_{ki}  +N\,\rho^{1-{\alpha}_{\min}}}  &\text{otherwise }
\end{cases}
\label{eq:AsympSINRRZFAprox}
\end{equation}

We now concentrate on the case where there exists at least one $\alpha_\ell \le 1$. We define 
\begin{equation}
\mu_{\min} \triangleq \min_{\substack{k,i\\k\neq i}} w_{ki} 
\label{eq:MinU}
\end{equation}
therefore in this special case we have:
\begin{align}
{\gamma}_k  &\dot{\leqslant}\frac{\big (\rho^{1-\alpha_{\min}}\big)^2  } { (N-1) \big(  \rho^{1-\alpha_{\min}} \big)^2 \mu_{\min} +N\,\rho^{1-\alpha_{\min}}}\label{eq:SINRRZFSecTerm} \\
&\doteq \frac{1}{(N-1)\mu_{\min}}\label{eq:SINRRZFPAsympUpper}\\
&\triangleq \bar{\gamma} \nonumber
\end{align}
Thus in general
\begin{align}
{\gamma}_k \;&\dot{\leqslant}\;\frac{\nu}{(N-1)\mu_{\min}} \label{eq:UBSINRRZF}\\
&\triangleq  \bar{\gamma} \nonumber
\end{align}
where $\nu$ is a new random variable defined as:
\begin{equation}
\nu =
 \begin{cases}
 \kappa_{\alpha} &\mbox{if $\alpha_k>1\; \forall k$ } \\
 1  &\mbox{otherwise}
\end{cases} \label{eq:CasesNu}
\end{equation}
where $\kappa_{\alpha}\triangleq \rho^{1-\alpha_{\min}}$.

We can now bound the outage probability as follows
\begin{align}
P_{\text{out}} &=\prob\bigg( \sum_{k=1}^{N}\log (1+\gamma_{k}) \leqslant R\bigg) \nonumber \\ &\,\dot{\geqslant} \;
 \prob\bigg( \sum_{k=1}^{N}\log (1+\bar{\gamma}) \leqslant R  \bigg)\nonumber\\
 &=  \prob\bigg(  \frac{\nu}{(N-1)\mu_{\min}} \leqslant 2^{R/N}-1  \bigg) \nonumber\\
&=  \prob\bigg(  \frac{\nu}{\mu_{\min}} \leqslant \Theta   \bigg) \label{eq:BoundRZFdiv}
\end{align}
where $\Theta \triangleq (2^{R/N}-1)(N-1)$.

The bound in~\eqref{eq:BoundRZFdiv} can be evaluated as follows
\begin{align}
  \prob\bigg( \frac{\nu}{\mu_{\min}} \leqslant \Theta \bigg)
&=\prob \big( \frac{\nu}{\mu_{\min}} \leqslant
  \Theta\big|\nu=\kappa_{\alpha} \big) \prob\big(\nu=\kappa_{\alpha} \big)+ \prob \big( \frac{\nu}{\mu_{\min}} \leqslant   \Theta\big|\nu=1 \big) \prob\big(\nu=1 \big)\nonumber\\
 &=\prob \big( \kappa_{\alpha} \leqslant
  \Theta \;\mu_{\min} \big) \prob\big(\nu=\kappa_{\alpha} \big)+ \prob \big( \frac{1}{\mu_{\min}} \leqslant   \Theta \big) \prob\big(\nu=1 \big).\label{eq:ProbRZF}
\end{align}
Notice that $\prob \big( \kappa_{\alpha} \leqslant \Theta \;\mu_{\min}
\big) \doteq 1$ since $\kappa_{\alpha}$ is vanishing at high SNR and $
\Theta $ and $ \mu_{\min} $ are positives.  We now need to compute
$\prob\big(\nu=\kappa_{\alpha} \big)$ and $\prob\big(\nu=1 \big)$, or
equivalently $\prob\big(\big\{\alpha_k>1\;\forall k \big\}\big)$ and
its complement. We quote one of the results of~\cite{Hesham:ISIT10}.
\begin{lemma}
\label{Lemma:AsympDist}
Let $\{\lambda_n \}$ denotes the eigenvalues of a Wishart matrix
$\bH\bH^H$, where $\bH$ is an $N \times M$ matrix with i.i.d Gaussian
entries, and let $\alpha_n=-\frac{\log(\lambda_n)}{\log(\rho)}$. If
${\bf{1}}_{\alpha_n}$ denotes the number of $\alpha_n$ that are
greater than one, then for any integer $s\leqslant N$ we
have~\cite[Section
  III-A]{Hesham:ISIT10}~\footnote{Note that~\cite{Hesham:ISIT10} analyzes
  linear MIMO receiver where it is assumed $N\geqslant M$. It can be
  easily shown that the above Lemma~\ref{Lemma:AsympDist} applies for
  the case considered here where $M\geqslant N$.}
\begin{equation}
\prob \big( {\bf{1}}_{\alpha_n} = s \big) \doteq \rho^{-(s^2+(M-N)s)}. \label{eq:CP_MIMO2}
\end{equation}

\end{lemma}

Thus setting $s=N$ (i.e. all $\alpha_n>1$) in~\eqref{eq:CP_MIMO2} yields
\begin{align}
\prob\big(\nu=\kappa_{\alpha} \big)&= \prob\big( {\bf{1}}_{\alpha_n} = N\big)\doteq \rho^{-MN}\label{eq:AsympValueAlpha1}
\\ \prob\big(\nu=1 \big)&\doteq O(1)\label{eq:AsympValueAlpha2}
\end{align}
where $O(1)$ is a non-zero constant with respect to $\rho$.

Evaluating~\eqref{eq:ProbRZF} depends on the values of $\Theta$ which
is always real and positive. If $\Theta < 1$ then we have
\begin{equation}
\prob\bigg( \frac{\nu}{\mu_{\min}} \leqslant \Theta \bigg)\doteq\rho^{-MN} \label{eq:ProbFullDiv}
\end{equation}
because $\prob \big( \frac{1}{\mu_{\min}} \leqslant   \Theta \big) =0$ as $1/\mu_{\min}>1$.
On the other hand if $\Theta>1$ then
\begin{align}
\prob\bigg( \frac{\nu}{\mu_{\min}} \leqslant \Theta \bigg)&\doteq\rho^{-MN}+ \prob \big( \frac{1}{\mu_{\min}} \leqslant   \Theta \big)  O(1) \label{eq:ProbDivZero1}\\
&\doteq O(1) \label{eq:ProbDivZero2}
\end{align}
since $\prob \big( \frac{1}{\mu} \leqslant \Theta \big)$ is not
a function of $\rho$ because $\mu$ is independent $\rho$. 
For the set of rates where $\Theta > 1$,
equation~\eqref{eq:ProbDivZero2} implies that the outage probability
in~\eqref{eq:BoundRZFend} is not function of $\rho$ and thus the
diversity is zero, i.e. the system will have error floor. The set of
rates for which $\Theta > 1$ are
\begin{equation}
R > N\log\big( \frac{N}{N-1}\big)\triangleq R_{th} .\label{eq:rateErroFloor}
\end{equation}


This concludes the calculation of a lower bound on the outage probability.
A similar approach will yield a corresponding upper bound, as follows. Let
\begin{equation}
\mu_{\max} \triangleq \max_{k\neq i} |u_{kl'} u^*_{il'}|^2 
\label{eq:MaxU}
\end{equation}
A lower bound on the SINR is given as
\begin{align}
{\gamma}_k \;&\dot{\geqslant}\;\frac{\nu}{(N-1)\mu_{\max}} \label{eq:LBSINRRZF}\\
&\triangleq \hat{\gamma}.\nonumber
\end{align}
The outage probability is bounded as
\begin{align}
P_{\text{out}} &\,\dot{\leqslant} \;
 \prob\bigg( \sum_{k=1}^{N}\log (1+\hat{\gamma}) \leqslant R  \bigg)\nonumber\\
&=  \prob\bigg(  \frac{\nu}{\mu_{\max}} \leqslant \Theta   \bigg). \label{eq:BoundRZFdiv2}
\end{align}

We can evaluate~\eqref{eq:BoundRZFdiv2} in a similar way
as~\eqref{eq:ProbRZF}, establishing that the outage diversity
$d_{out}^{RZF}=MN$ if the operating spectral efficiency $R$ is less
than $R_{th}=N\log{(\frac{N}{N-1})}$, and $d_{out}^{RZF}=0$ if
$R>R_{th}$.  This shows that the performance of RZF precoder can be
much better than that of the conventional ZF precoder MIMO system
whose diversity is $M-N+1$ independent of rate.

Recall that diversity is the SNR exponent of the probability of
codeword error. In Appendix~\ref{appendix:PEP}, we show that the outage exponent
tightly bounds the SNR exponent of the error probability. Thus we have
the following theorem.

\begin{theorem}
\label{Th:RZFP}
For an $M\times N$ MIMO system that utilizes joint spatial encoding
and regularized ZF precoder given by~\eqref{eq:RZFPrecoderFilter}, the
outage diversity is $d^{RZF}=MN$ if the operating spectral efficiency
$R$ is less than $R_{th}=N\log{(\frac{N}{N-1})}$, and $d^{RZF}=0$ if
$R>R_{th}$ .
\end{theorem}

\begin{remark}

$R_{th}$ is a monotonically decreasing function of $N$ with the
  asymptotic value $\lim_{N\to\infty}R_{th}=\frac{1}{\ln 2}\approx
  1.44$. Overall we have $1.44 \le R_{th} \le 2$, leading to an easily
  remembered rule of thumb that applies to all antenna
  configurations. Regularized ZF precoders always exhibit an error
  floor at spectral efficiencies above $2$ b/s/Hz, and enjoy full
  diversity at spectral efficiencies below $1.44$ b/s/Hz.
\end{remark}

\subsection{Matched Filter Precoding}
\label{sec:MFPrec}

The transmit matched filter (TxMF) is introduced
in~\cite{Esmailzadeh:ICC93,Joham:JSP05}. The TxMF maximizes the
signal-to-interference ratio (SIR) at the receiver and is optimum for
high signal-to-noise-ratio scenarios~\cite{Joham:JSP05}. The TXMF is
also proposed for non-cooperative cellular wireless
network~\cite{Marzetta:TW10}. The TxMF is derived by maximizing the
ratio between the power of the desired signal portion in the received
signal and the signal power under the transmit power constraint, that
is~\cite{Joham:JSP05}
\begin{align}
\bP= &\arg\max_{\bP} \frac{E \big( ||\bx^H\tilde{\by}||^{2} \big)}{E\big(||\bn||^2\big)} \label{eq:MFOptim}\\
&\text{subject to:  }   E||\bP\bx||^2\leqslant \rho \nonumber
\end{align}
where $\tilde{\by}$ is the noiseless received signal $\tilde{\by} = \bP \bx$.

The solution to~\eqref{eq:MFOptim} is given by
\begin{equation}
\bP = \beta \bH^H \label{eq:MFPrecRes}
\end{equation}
with
\begin{equation}
\beta=\sqrt{\frac{1}{\text{tr}(\bH^H\bH)}}. \label{eq:ScalarMFPrec}
\end{equation}

We now analyze the diversity for the MIMO system under TxMF.
The received signal is given by
\begin{equation}
\by=\bH\bH^H\bx+\bn=\beta\bU\Lambda\bU^H\bx+\bn .\nonumber
\end{equation}
The received signal at the $k$-th antenna 
\begin{align}
y_k &= \beta \bigg( \sum_{l=1}^{N} \lambda_l |u_{kl}|^2 \bigg) x_k \;\;+\nonumber\\
 &\quad\; \beta\sum_{i=1,i\neq k}^{N} \bigg( \sum_{l=1}^{N} \lambda_l u_{kl}u^*_{il} \bigg)  x_{i}  + n_{k}\label{eq:RecSigMFP}
\end{align}
The SINR at $k$-th receive antenna is
\begin{equation}
\gamma_{k}=\frac{\beta^2\frac{\rho}{N} \bigg( \sum_{l=1}^{N} \lambda_l |u_{kl}|^2 \bigg)^2  }{\beta^2\frac{\rho}{N} \sum_{i=1,i\neq k}^{N} \bigg| \sum_{l=1}^{N} \lambda_l u_{kl}u^*_{il} \bigg|^2 + 1} \nonumber
\end{equation}

Substitute with the value of $\beta$ and $\lambda_l=\rho^{-\alpha_l}$
\begin{align}
\gamma_k&= \frac{\bigg( \sum_{l=1}^{N} \rho^{-\alpha_l}  |u_{kl}|^2 \bigg)^2 } {\sum_{i=1,i\neq k}^{N} \bigg| \sum_{l=1}^{N} \rho^{-\alpha_l} u_{kl} u^*_{il} \bigg|^2 +N\,\rho^{-1} \sum_{l=1}^N \rho^{-\alpha_l}} \label{eq:SINRMFPAproxim1}
\end{align}
Observe that~\eqref{eq:SINRMFPAproxim1} is the same as the SINR of the
RZF precoded system given by~\eqref{eq:SINRRZFDef2}. Hence the
analysis in the present case follows closely that of the outage lower
bound of the RZF precoder, with the following result: the system can
achieve full diversity as long as the operating rate is less than
$R_{th}$ given in~\eqref{eq:rateErroFloor}.  The pairwise error
probability analysis is also similar to that of the RZF precoding
system (given in Appendix~\ref{appendix:PEP}) which we omit for
brevity. Thus we conclude that Theorem~\ref{Th:RZFP} applies for the
TxMF precoder.

\subsection{Wiener Filter Precoding}
\label{sec:WFPrec}

The transmit Wiener filter TxWF minimizes the weighted MSE
function. 
\begin{align}
\{ \bT , \beta\} = &\text{argmin}_{\bT,\beta} E\big( ||\bx-\beta^{-1}\tilde{\by} ||^2  \big) \nonumber\\
 &\text{subject to } E \big( ||\bT\bx  \big ||^2) \leqslant \rho. \label{eq:EFProblem}
\end{align}

Solving~\eqref{eq:EFProblem} yields
\begin{equation}
\bT = \beta \bF^{-1} \bH^H \label{eq:WFPrecoder}
\end{equation}
with
\begin{align}
\bF&= \bigg(\bH^H\bH + \frac{N}{\rho}\bI\bigg)\nonumber\\
\beta &= \sqrt{\frac{1}{\text{tr}(\bF^{-2}\bH^H\bH)}} \label{eq:ScalarWFP}
\end{align}
where $\beta$ can be interpreted as the optimum gain for the
combined precoder and channel~\cite{Joham:JSP05}. 

Notice that the TxWF precoding function is similar to that of the MMSE
equalizer~\cite{Verdu98:book}. Indeed the SINR of both systems are
equivalent. To see this, we first compute the SINR for the precoded
$\bH\in\mathbb{C}^{M\times N}$ (with $M\geqslant N$) MIMO channel
\begin{align}
\gamma_{k}&=\frac{\frac{\rho\,\beta}{N}|(\bP\;\bH)_{kk}|^2}{\frac{\rho\,\beta}{N}\sum_{i\neq k}^N |(\bP\;\bH)_{ki}|^2 + 1}\label{eq:SINRWFIndSig}\\
&=\frac{\frac{\rho}{N}|(\bP\;\bH)_{kk}|^2}{\frac{\rho}{N}\sum_{i\neq k}^N |(\bP\;\bH)_{ki}|^2 + \text{tr}(\bF^{-2}\bH^H\bH)}\label{eq:SINRWFP}
\end{align}
where we have used the independence of the transmitted signal to
compute~\eqref{eq:SINRWFIndSig}.

Now consider a MIMO channel $\bH_2=\bH^T \in \mathbb{C}^{N \times M}$. The MMSE equalizer for this channel is
given by
\begin{equation}
\bW_{e} =  (\bH_2^H\bH_2+\frac{N}{\rho}\bI)^{-1} \bH_2^H .\label{eq:MMSEEq}
\end{equation}

The received SINR for that system is given by
\begin{equation}
\gamma^{MMSE}_{k}=\frac{\frac{\rho}{N}|(\bW_{e}\;\bH_2)_{kk}|^2}{\frac{\rho}{N}\sum_{i\neq k}^N |(\bW_{e}\;\bH_2)_{ki}|^2 + \text{tr}(\bW_{e}\bW_{e})}.\label{eq:SINRMMSEEq}
\end{equation}

Since $\bW_{e}\;\bH_2$ = $\bP_{WFP}\bH$ and $\text{tr}(\bW_{e}\bW_{e})$
= $ \text{tr}(\bF^{-2}\bH^H\bH)$, we conclude that $\gamma^{MMSE}_{k} = \gamma^{WFP}_{k}$. Hence the diversity analysis
of~\cite{Hesham:ISIT10,Kumar:JIT09} for the MIMO MMSE receiver applies
for the MIMO Wiener precoding system. It is shown
in~\cite{Hesham:ISIT10} that this diversity is a function of rate $R$ and number of transmit and receive antennas. We thus conclude the following.

\begin{lemma}
Consider a channel $\bH\in\mathbb{C}^{M\times N}$ the diversity of
the MIMO system under Wiener filter
precoding is given by
\begin{equation}
d^{WFP}= \lceil N2^{-\frac{R}{N}} \rceil^{2}+ (M-N)\lceil N
2^{-\frac{R}{N}}\rceil \label{eq:DivMMSEAndWFP}
\end{equation}
 where $(\cdot)^+ = \max(\cdot,0)$ and $\lceil \cdot \rceil$. 
\end{lemma}

\begin{remark}
It is commonly stated that MMSE and ZF operators ``converge'' at high
SNR. The developments in this paper as well as~\cite{Hedayat:JSP07}
serve to show that although not false, this comment is essentially
fruitless because the performance of MMSE and ZF at high SNR are very
different. This apparent incongruity is explained in the broadest
sense as follows: Even though the MMSE coefficients converge to ZF
coefficients as $\rho\rightarrow\infty$, the high sensitivity of
logarithm of errors (especially at low error probabilities) to
coefficients is such that the convergence of MMSE to ZF coefficients
is not fast enough for the logarithm of respective errors to converge.
\end{remark}

\section{Diversity-Multiplexing Tradeoff in Precoding}
\label{sec:DMTPrecOnly}

For increasing sequence of SNRs, consider a corresponding sequence of
codebooks $\mathcal{C}(\rho)$, designed at increasing rates $R(\rho)$
and yielding average error probabilities $P_e(\rho)$. Then define
\begin{align}
r &=\lim_{\rho\to\infty} \frac{R(\rho)}{\log{\rho}}
\nonumber\\ d&=-\lim_{\rho\to\infty} \frac{\log
  P_e(\rho)}{\log{\rho}} . \nonumber
\end{align}
For each $r$ the corresponding diversity $d(r)$ is defined (with a
slight abuse of notation) as the supremum of the diversities over all
possible codebook sequences $\mathcal{C}(\rho)$.

From the viewpoint of definitions, the traditional notion of diversity
can be considered a special case of the DMT by setting $r=0$. However,
from the viewpoint of analysis, the approximations needed in DMT
calculation make use of $R(\rho)$ being a {\em strictly} increasing
function, while for diversity analysis $R$ is constant (not strictly
increasing function of $\rho$). Thus, although sometimes DMT analysis
may produce results that are luckily consistent with diversity
analysis\footnote{E.g. the point-to-point MIMO channel with ML
  decoding.} ($r=0$), in other cases one may not be so lucky and the
DMT analysis may produce results that are inconsistent with diversity
analysis. Certain equalizers and precoders fall into the latter
category. In the following, we calculate the DMT of the various
precoders considered up to this point.

\subsubsection{ZF Precoding}
\label{sec:DMTZF}
Recall that two ZF precoding designs have been considered.  For the ZF
precoder minimizing power, given by~\eqref{eq:ZFFirstDesignTotPower},
the outage upper bound in~\eqref{eq:OutProbUBZFFirstDes} can be
written as
\begin{align}
P_{out} &\leqslant \prob \big( \lambda_{\min} \leqslant \rho^{(\frac{r}{N}-1)} \big) \label{eq:BoundDMTZFPrec1}\\
&\doteq \rho^{-(M-N+1)(1-\frac{r}{N})} \label{eq:BoundDMTZFPrec2}
\end{align}
where we substitute $R= r\log\rho$ to
obtain~\eqref{eq:BoundDMTZFPrec1}, and
equation~\eqref{eq:BoundDMTZFPrec2} follows in a manner identical to
the procedure that led to~\eqref{eq:JointEncZFUBResult}.

Similarly the outage lower bound~\eqref{eq:JointEncOutZFLB2} can be written as 
\begin{align}
P_{out} &\geqslant \prob \big( z \leqslant \rho^{(\frac{r}{N}-1)} \big) \nonumber\\
&\doteq \rho^{-(M-N+1)(1-\frac{r}{N})}. \label{eq:BoundDMTZFPrec3}
\end{align}

From~\eqref{eq:BoundDMTZFPrec2} and~\eqref{eq:BoundDMTZFPrec3} we conclude
\begin{equation}
d^{ZFP}(r)= (M-N+1) \big( 1-\frac{r}{N}  \big)^+. \label{eq:DMTZFPrec}
\end{equation}

The DMT of the ZF precoder maximizing the throughput, given
by~\eqref{eq:ZFPrecThroOptimal}, is obtained in an essentially similar
manner to the above, therefore the discussion is omitted in the
interest of brevity.

\subsubsection{Regularized ZF Precoding}
\label{sec:DMTRZF}
We begin by producing an outage lower bound. To do so, we start by the
bound on the SINR of each stream $k$ obtained
in~\eqref{eq:AsympSINRRZFAprox}, and further bound it by discarding
some positive terms in the denominator.
\begin{align}
&\bar\gamma_k=\frac{\big (\rho^{1-\alpha_{\min}}\big)^2  } { \sum_{\substack{i\neq k}} \big|  u_{kl} u^*_{il}  \,\rho^{1-\alpha_{\min}}   \big|^2 +N\rho^{1-\alpha_{\min}}}\nonumber\\
&\qquad\qquad\leqslant \begin{cases}
\frac{\big (\rho^{1-\alpha_{\min}}\big)^2 }{\rho^{2(1-\alpha_{\min})}\big|u_{kl'}u^*_{2l'}\big|^2+N\rho^{1-\alpha_{\min}}} & k=1\\
\frac{\big (\rho^{1-\alpha_{\min}}\big)^2 }{\rho^{2(1-\alpha_{\min})}\big|u_{kl'}u^*_{1l'}\big|^2+N\rho^{1-\alpha_{\min}}}& k>1
\end{cases}\nonumber\\
&\qquad\qquad\doteq  \begin{cases}
\frac{1} {|u_{kl'} u^*_{2l'}|^2} & k=1 \\
\frac{1} {|u_{kl'} u^*_{1l'}|^2} & k>1
\end{cases}\nonumber
\end{align}

We can now bound the outage probability   
\begin{align}
P_{\text{out}} &=\prob\bigg( \sum_{k=1}^{N}\log (1+\gamma_k) \leqslant R  \bigg) \nonumber \\ 
&\,\dot{\geqslant} \;
 \prob\bigg( \sum_{k=1}^{N}\log (1+\bar{\gamma}_k) \leqslant R  \bigg)\nonumber\\
 &\geqslant \prob\bigg( N\log \sum_{k=1}^{N}\frac{1}{NM_{s}}(1+\bar{\gamma}_k) \leqslant R  \bigg)\label{eq:SpechtBound}\\
 &\doteq \prob\bigg(\sum_{k=1}^{N}\frac{1}{NM_{s}}(1+\bar{\gamma}_k) \leqslant \rho^{\frac{r}{N}}  \bigg)\label{eq:SpechtRemove}\\
 &\doteq \prob\bigg(\sum_{k=1}^{N}\frac{1}{N}(1+\bar{\gamma}_k) \leqslant \rho^{\frac{r}{N}}  \bigg)\label{eq:SpechtRemove2}\\
 &\doteq \prob\bigg(\sum_{k=1}^{N}\bar{\gamma}_k \leqslant \rho^{\frac{r}{N}}  \bigg)\nonumber\\
&\; \dot{\geqslant}  \; \prob\bigg( \frac{\nu}{|u_{kl'} u^*_{2l'}|^2}+\sum_{k=2}^{N}\frac{\nu}{|u_{kl'} u^*_{1l'}|^2} \leqslant \rho^{\frac{r}{N}}  \bigg). \label{eq:BoundRZFend}
\end{align}
where we have used the Specht bound in~\eqref{eq:SpechtBound} in a
manner similar
to~\cite{Hesham:ISIT10}. Equation~\eqref{eq:SpechtRemove}
and~\eqref{eq:SpechtRemove2} follow similarly to~\cite[Section
  III-B]{Hesham:ISIT10}

For notational convenience define 
\[
\psi\defeq\frac{1}{|u_{kl'}
  u^*_{2l'}|^2}+\sum_{k=2}^{N}\frac{1}{|u_{kl'} u^*_{1l'}|^2} \; .
\]
Then
the bound in~\eqref{eq:BoundRZFend} can be evaluated as follows:
\begin{align}
  &\prob\bigg( \nu\psi \leqslant \rho^{\frac{r}{N}} \bigg) \nonumber\\
&=\prob \big( \nu\psi \leqslant
  \rho^{\frac{r}{N}}\big|\nu=0 \big) \prob\big(\nu=0 \big)+ \prob \big( \nu\psi \leqslant   \rho^{\frac{r}{N}}\big|\nu=1 \big) \prob\big(\nu=1 \big)\nonumber\\
 &=\prob \big( 0 \leqslant
  \rho^{\frac{r}{N}} \big) \prob\big(\nu=0 \big)+ \prob \big( \psi \leqslant   \rho^{\frac{r}{N}} \big) \prob\big(\nu=1 \big)\nonumber\\
& \doteq \rho^{-MN} +  \prob \big(\psi \leqslant   \rho^{\frac{r}{N}}  \big) \;O(1).\label{eq:ProbRZFDMT}\\
&\geqslant \rho^{-MN} +  O(1) \label{eq:ProbRZFDMT2}\\
&=O(1)
\end{align}
where~\eqref{eq:ProbRZFDMT} follows from Lemma~\ref{Lemma:AsympDist},
and~\eqref{eq:ProbRZFDMT2} is true as long as $ \prob
\big(\psi \leqslant
\rho^{\frac{r}{N}} \big) = O(1)$, the proof of which is relegated to 
Appendix~\ref{appendix:A}.

Since the outage lower bound~\eqref{eq:ProbRZFDMT2} is not a function of
$\rho$, the system will always have an error floor. In other words the
DMT is given by
\begin{align}
d^{RZFP}(r) = 0 \qquad 0 < r \le B \label{eq:DMTRZFPEndRes}
\end{align}

We saw earlier that in the fixed-rate regime RZF precoding enjoys full
diversity for spectral efficiencies below a certain threshold, but it
now appears that DMT shows only zero diversity. DMT is not capable of
predicting the complex behavior at $r=0$ because the DMT framework
only assigns a single value diversity to all distinct spectral
efficiencies at $r=0$.  A similar behavior was observed and analyzed
for the MMSE MIMO receiver~\cite{Hedayat:JSP07,Kumar:JIT09,
  Hesham:ISIT10}.

\subsubsection{Matched Filter Precoding}
\label{sec:DMTMF}

The DMT of the MIMO system with TxMF is the same as the DMT given
by~\eqref{eq:DMTRZFPEndRes} due to the similarity in the outage
analysis (see Section~\ref{sec:MFPrec}). We omit the details for brevity.

\subsubsection{Wiener Filter Precoding}
\label{sec:DMTWF}

Since the the received SINR of the MIMO system using TxWF precoding is the same as that of MIMO MMSE receiver, we conclude from~\cite{Kumar:JIT09} that the DMT for the TxWF precoding system is  
\begin{equation}
d^{WFP}(r)= (M-N+1) \big( 1-\frac{r}{N}  \big)^+. \label{eq:DMTWFPrec}
\end{equation}

Similarly to the MIMO MMSE receiver~\cite{Kumar:JIT09,Hesham:ISIT10},
we observe that DMT for the MIMO system with TxWF does not always
predict the diversity in the fixed rate regime given
by~\eqref{eq:DivMMSEAndWFP}.

\section{Equalization for Linearly Precoded Transmission}
\label{sec:JointLinearTxRx}

\begin{figure*}
\centering
\includegraphics{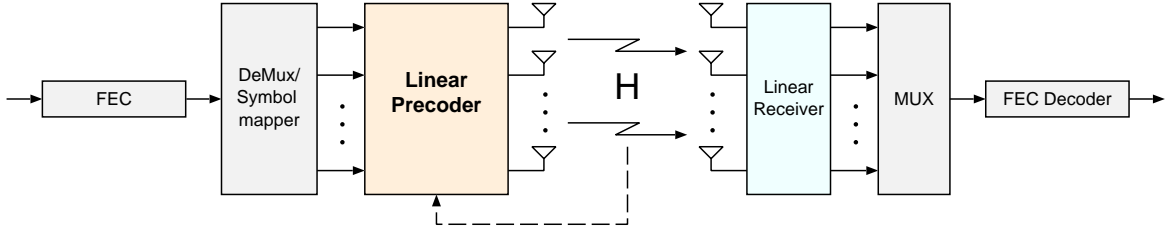}
\caption{MIMO with linear precoder with receive-side equalization}
\label{fig:linearprecoderEq}
\end{figure*}

The objective of a precoded transmitter is to separate the data
streams at the receiver. In other words, linear precoding is a method
of interference management at the transmitter. In general, precoded
systems do not require interference management at the receiver,
however, once a transmitter is designed and standardized (as precoders
have been), some standards-compliant receivers may opt to further
equalize the precoded channel (see
Figure~\ref{fig:linearprecoderEq}). This section analyzes the
equalization of precoded transmissions.

When the transmit and receive filters can be designed jointly and from
scratch, singular value decomposition becomes an attractive option
whose diversity has been analyzed in~\cite{Ersin:Comm06}. The
distinction of the systems analyzed in this section is that the
precoders can be used with or without the receive filters, while with
the SVD solution neither the transmit nor the receive filters can
operate without each other.

A snapshot of some of the results of this section is as follows. It is
shown that equalization at the receiver can alleviate the error floor
that was observed in matched filter precoding as well as regularized
ZF precoding. It is shown that MMSE equalization does not affect the
diversity of Wiener filter precoding, but ZF equalization does indeed
affect the diversity of Wiener filter precoding in a negative way.

Recall that in the system model given in Section~\ref{sec:SysMod} we
have defined the precoder and equalizer matrices $\bP \in
\mathbb{C}^{M\times B}$ and $\bW \in\mathbb{C}^{B\times N}$,
respectively, where $B$ is the number of information symbols not to
exceed $\min (M,N)$. In most wireless systems, the equalizer at the
receiver is designed to equalize the compound channel ($\bH\bP$)
composed of the precoder and the channel (rather than designing the
precoder for the equalized channel ($\bW\bH$) although it is
possible). In such case we have $M \geqslant N$ and we set $B=N$.

\subsection{ZF Equalizer}
\label{sec:ZFEq}

The ZF equalizer is analyzed when operating together with various precoders, as follows.

\subsubsection{Wiener Filter Precoding}
\label{section:WienerZF}

The TxWF precoder is given by
\begin{align}
\bT &= \beta \bigg(\bH^H\bH + \frac{N}{\rho}\bI\bigg)^{-1} \bH^H \nonumber\\
&= \beta \bH^H \bigg(\bH\bH^H + \frac{N}{\rho}\bI_N\bigg)^{-1}  \label{eq:WFPrecoder2}
\end{align}
where~\eqref{eq:WFPrecoder2} follows from~\cite[Fact
  2.16.16]{Bernstein09:book}~\footnote{Let $\bA\in \mathbb{C}^{n\times
    m}$ and $\bB \in \mathbb{C}^{m\times n}$ then
  $(\bI_n+\bA\bB)^{-1}\bA=\bA(\bI_m+\bB\bA)^{-1}$. This fact can be
  proved via Matrix Inversion Lemma.}.  The scalar coefficient $\beta$
is given in~\eqref{eq:ScalarWFP} and, similar
to~\eqref{eq:NormFactor}, it can be written as $\beta =1/\sqrt{\eta}$
\begin{equation}
\eta=\text{tr}\big[\Lambda(\Lambda+N\rho^{-1}\,\bI)^{-2}\big]=\sum_{l=1}^{N}\frac{\lambda_l}{(\lambda_l+N\rho^{-1}\,)^2} \nonumber
\end{equation}

The ZF equalizer for the precoder and the channel is given by
\begin{align}
\bW_{ZF} &= (\mathbb{H}^H\mathbb{H})^{-1}\mathbb{H}^H \label{eq:ZFEqChWF}
\end{align}

The composite channel $\mathbb{H}$ is given by
\begin{equation}
\mathbb{H} = \bH \bT. \nonumber
\end{equation}
 
The received signal is given by
\begin{equation}
 \by = \bW_{ZF}\bH \bP\bx + \bW_{ZF}\bn. \label{eq:SysModWFPZFEq}
\end{equation}

The filtered noise $\tilde{\bn}=\bW_{ZF}\bn$ is is a complex Gaussian
vector with zero-mean and covariance matrix $R_{\tilde{n}}$ given by
\begin{align}
R_{\tilde{n}} &= [\mathbb{H}^H\mathbb{H}]^{-1} \nonumber\\
&=\big[(\bH\bH^H+N\rho^{-1}\,\bI)^{-1}(\bH\bH^H)^2 (\bH\bH^H+N\rho^{-1}\,\bI)^{-1}\big]^{-1}\nonumber\\
&=\big[\bU\Lambda(\Lambda+N\rho^{-1}\,\bI)^{-1}\bU^H\bU\Lambda(\Lambda+N\rho^{-1}\,\bI)^{-1}\bU^H\big]^{-1}\nonumber\\
&=\big[\bU\Lambda^2(\Lambda+N\rho^{-1}\,\bI)^{-2}\bU^H\big]^{-1}\nonumber
\end{align}
where we have used the eigen decomposition $\bH\bH^H=\bU\Lambda\bU^H$. The noise variance of the output stream $k$ is therefore
\begin{equation}
R_{\tilde{n}}(k,k) = \sum_{l=1}^{N} \bigg(\frac{\lambda_l+N\rho^{-1}}{\lambda_l}\bigg)^2 |u_{kl}|^2  \label{eq:GetNoiseCov}
\end{equation}
where~\eqref{eq:GetNoiseCov} follows in a similar manner
as~\eqref{eq:GetInvEigDecomp}. 
We can compute the signal-to-noise ratio of the ZF filter output:
\begin{align}
\gamma_{k} &= \frac{\rho \;\beta^2}{N\,R_{\tilde{n}}(k,k)}\nonumber\\
 &= \frac{\rho/N}{  \sum_{j=1}^{N}\frac{\lambda_j}{(\lambda_j+N\rho^{-1}\,)^2}      \sum_{l=1}^{N} \big(\frac{\lambda_l+N\rho^{-1}}{\lambda_l}\big)^2 |u_{kl}|^2 }\label{eq:SINRWFPZFEq}.
\end{align}


Due to the complexity of~\eqref{eq:SINRWFPZFEq} we proceed to bound the outage from above and below.
The upper bound on outage is calculated as follows. Since $|u_{kl}|
\leqslant 1$, 
\begin{align}
\gamma_{k} &\geqslant \frac{\rho/N}{  \sum_{j=1}^{N}\frac{\lambda_j}{(\lambda_j+N\rho^{-1}\,)^2}      \sum_{l=1}^{N} \big(\frac{\lambda_l+N\rho^{-1}}{\lambda_l}\big)^2  }\label{eq:BoundSINRWFPZFEq}\\
  &= \frac{1/N}{  \sum_{j=1}^{N}\frac{\rho^{1-\alpha_j}}{(\rho^{1-\alpha_j}+N\,)^2}      \sum_{l=1}^{N} \big(\frac{\rho^{1-\alpha_l}+N}{\rho^{1-\alpha_l}}\big)^2  } \label{eq:SNIRGeneralEx}\\
&\triangleq \hat{\gamma} .\label{eq:BoundSINRWFPZFEq2}
\end{align}
where we have substituted $\lambda_l=\rho^{-\alpha_l}$
in~\eqref{eq:SNIRGeneralEx}.  Thus the outage probability is bounded
as
\begin{align}
P_{\text{out}}&= \prob\bigg( \sum_{k=1}^{N}\log (1+\gamma_{k}) \leqslant R\bigg) \nonumber\\
&\leqslant  \prob\bigg( \sum_{k=1}^{N}\log (1+\hat{\gamma}) \leqslant R\bigg) \nonumber\\
&= \prob\bigg(\hat{\gamma} \leqslant 2^{\frac{R}{N}} -1\bigg) \label{eq:OutProbWFPZFEqu}
\end{align}
Similarly to the previous analysis, we examine the SINR bound
$\hat{\gamma}$ for different values of $\alpha_l$. Define the set ${\mathcal B} = \{ l \; | \; \alpha_l > 1\}$ and the event
\begin{equation}
\mathcal{L} = \{ |{\mathcal B} | =N \} \label{eq:EventAlphas}
\end{equation}
 we have
\begin{align}
P_{\text{out}} &\leqslant  \prob\bigg(\hat{\gamma} \leqslant 2^{\frac{R}{N}} -1\bigg) \nonumber\\
&= \prob\bigg(\hat{\gamma} \leqslant 2^{\frac{R}{N}} -1  \big| \mathcal{L}  \bigg) \prob(\mathcal{L}) + \prob\bigg(\hat{\gamma} \leqslant 2^{\frac{R}{N}} -1  \big| \bar{\mathcal{L}}  \bigg) \prob(\bar{\mathcal{L}}) \label{eq:OutUNEFPZFEAA}\\
&\leqslant \prob\bigg(\hat{\gamma} \leqslant 2^{\frac{R}{N}} -1  \big| \mathcal{L}  \bigg) +  \prob\bigg(\hat{\gamma} \leqslant 2^{\frac{R}{N}} -1  \big| \bar{\mathcal{L}}  \bigg) . \label{eq:OutUNEFPZFE}
\end{align}

To calculate the first term in~\eqref{eq:OutUNEFPZFE}, we evaluate
$\hat{\gamma}$ when $\alpha_l \geqslant 1 \;\forall l$
\begin{align}
   \hat{\gamma} &\doteq \frac{1/N}{ \sum_{j=1}^{N}\rho^{1-\alpha_j}      \sum_{l=1}^{N} \frac{1}{\rho^{2(1-\alpha_l)}}  }  \label{eq:SINRWFPZEAsympAprox2}\\
&\,\dot{\geqslant}\,  \frac{1/N}{ \sum_{l=1}^{N} \frac{1}{\rho^{2(1-\alpha_l)}}  } \label{eq:SINRWFPZEAsympAprox4}\\
&\doteq  \frac{1}{N} \rho^{2(1-\alpha_{\max})} =  \frac{1}{N} \rho^2 \lambda_{\min}^2 \label{eq:SINRWFPZEAsympAprox3}
\end{align}
where~\eqref{eq:SINRWFPZEAsympAprox2} follows because
$\rho^{1-\alpha_l}+N \doteq N$,~\eqref{eq:SINRWFPZEAsympAprox4}
follows because $\sum_{j=1}^{N}\rho^{1-\alpha_j} \dot{\leqslant} 1$,
and~\eqref{eq:SINRWFPZEAsympAprox3} follows because the sum
in~\eqref{eq:SINRWFPZEAsympAprox4} is asymptotically dominated by the
largest component.

We continue to bound the first term in~\eqref{eq:OutUNEFPZFE} 
\begin{align}
\prob\bigg(\hat{\gamma} \leqslant 2^{\frac{R}{N}} -1  \big| \mathcal{L}  \bigg) &\dot{\leqslant} \; \prob \bigg(  \frac{1}{N} \rho^2 \lambda_{\min}^2 \leqslant 2^{\frac{R}{N}} \bigg) \nonumber\\
&\doteq \prob \bigg( \lambda_{\min} \leqslant \rho^{-1} \bigg) \label{eq:WFPZFUBOutage}\\
&\doteq \rho^{-(M-N+1)}\label{eq:WFPZFUBOutage2}
\end{align}
where~\eqref{eq:WFPZFUBOutage} is the same
as~\eqref{eq:ZFPrecOutLBLambdaMin} , hence~\eqref{eq:WFPZFUBOutage2} follows.

 To calculate the second term in~\eqref{eq:OutUNEFPZFE}, we evaluate
 $\hat{\gamma}$ when one or more $\alpha_l \leqslant 1$. Consider the
 the two summations in the denominator
 of~\eqref{eq:SNIRGeneralEx}. The first one can be asymptotically
 evaluated as
\begin{align}
 &\sum_{j=1}^{N}\frac{\rho^{1-\alpha_j}}{(\rho^{1-\alpha_j}+N\,)^2} \doteq  \sum_{\alpha_j<1}\frac{1}{\rho^{1-\alpha_j}} +  \sum_{\alpha_j>1}\rho^{1-\alpha_j} \nonumber\\
&\doteq\begin{cases}
 \rho^{-(1-\alpha_{\max})} &\text{$|\bar{\mathcal{L}}|=N$ } \\
\max(\rho^{-1+\alpha'},\rho^{1-\alpha''})\;\dot{\leqslant}\; \rho^{-(1-\alpha_{\max})} &\text{$1\leqslant |\bar{\mathcal{L}}| < N$}
\end{cases}\label{eq:EqAsympSINREig}
\end{align}
where $\alpha'=\max_{\alpha_j<1}\alpha_j$ and
$\alpha''=\min_{\alpha_j>1}\alpha_j$ and~\eqref{eq:EqAsympSINREig}
follows because
$\min(\rho^{-1+\alpha'},\rho^{1-\alpha''})\;\dot{\leqslant}\;\rho^{-(1-\alpha_{\max})}$.
The second summation in the denominator of~\eqref{eq:SNIRGeneralEx} can be evaluated as follows
\begin{align}
&\sum_{l=1}^{N} \bigg(\frac{\rho^{1-\alpha_l}+N}{\rho^{1-\alpha_l}}\bigg)^2 \doteq \sum_{\alpha_l < 1 } 1 + \sum_{\alpha_l > 1} \frac{1}{\rho^{2(1-\alpha_l)}} \nonumber\\
&\doteq  \begin{cases}
1 &\text{$|\bar{\mathcal{L}}|=N$ } \\
\rho^{-2(1-\alpha_{\max})} &\text{$1\leqslant |\bar{\mathcal{L}}| < N$}
\end{cases}\label{eq:SecondSumAsymptot}
\end{align}

We now use~\eqref{eq:EqAsympSINREig} and~\eqref{eq:SecondSumAsymptot} to
bound $\hat{\gamma}$
\begin{align}
\hat{\gamma}&\dot{\geqslant} \begin{cases}
  {\rho^{1-\alpha_{\max}}}\,\,=\rho\lambda_{\min} &|\bar{\mathcal{L}}|=N
   \\ {\rho^{2-2\alpha_{\max}}}=\rho^3\lambda^3_{\min}  &1\leqslant
    |\bar{\mathcal{L}}| < N
\end{cases} 
\nonumber\\
&\triangleq \bar{\gamma}
\end{align}
 We thus have
\begin{align}
&P_{out} \leqslant \prob\bigg(\hat{\gamma}\leqslant 2^{\frac{R}{N}} -1  \bigg| \bar{\mathcal{L}}  \bigg)\nonumber\\ &\leqslant \prob\bigg(\bar{\gamma}\leqslant 2^{\frac{R}{N}} -1  \bigg| \bar{\mathcal{L}}  \bigg) \nonumber\\
&< \prob\bigg(\bar{\gamma} \leqslant 2^{\frac{R}{N}} -1  \;\bigg|\;  |\mathcal{B}|=0   \bigg) +\prob\bigg(\bar{\gamma} \leqslant 2^{\frac{R}{N}} -1  \;\bigg|\;  0<
    |\mathcal{B}| < N  \bigg) \nonumber\\ 
&\doteq \prob \big( \lambda_{\min} \leqslant \rho^{-1}  \big)+\prob \big( \lambda^3_{\min} \leqslant \rho^{-3}  \big) \nonumber\\
&\doteq \prob \big( \lambda_{\min} \leqslant \rho^{-1}  \big) \nonumber\\
&\doteq \rho^{-(M-N+1)}. \label{eq:OutBoundAlphaGreaOne2}
\end{align}

This concludes the calculation of outage upper bound. We now proceed with the outage lower bound.


Define the event $\mathcal{P}= \{ |a_{kl}| \geqslant \epsilon \quad
\forall \, k,l \}$ where $a_{kl}$ is the $(k,l)$ entry of the unitary
matrix $U$ (c.f. equation~\eqref{eq:GetInvEigDecomp}). Define
\begin{align}
\breve{\gamma} &= \frac{1/N}{  \sum_{j=1}^{N}\frac{\rho^{1-\alpha_j}}{(\rho^{1-\alpha_j}+N\,)^2}      \sum_{l=1}^{N} \big(\frac{\rho^{1-\alpha_l}+N}{\rho^{1-\alpha_l}}\big)^2 \epsilon }\label{eq:SINRTwoSum}
\end{align}
Notice that $\breve{\gamma} > \gamma$ because
$|a_{kl}| \geqslant \epsilon \quad \forall \, k,l$.

The outage probability is bounded as
\begin{align}
P_{\text{out}}&= \prob\bigg( \sum_{k=1}^{N}\log (1+\gamma_{k}) \leqslant R\bigg) \nonumber\\
&\geqslant  \prob\bigg( \sum_{k=1}^{N}\log (1+\gamma_{k}) \leqslant R \bigg| \mathcal{P} \bigg) \; \prob ({\mathcal{P}}) \nonumber \\
&\geqslant  \prob\bigg( \sum_{k=1}^{N}\log (1+\breve{\gamma}) \leqslant R  \bigg) \; \prob ({\mathcal{P}}) \label{eq:OutLBZFEqWFP1}\\
&=  \prob\bigg( \breve{\gamma} \leqslant 2^{\frac{R}{N}}-1  \bigg) \; \prob ({\mathcal{P}}) \label{eq:OutLBWFPZF2}
\end{align}

The probability
$\prob (\mathcal{P}) = O(1)$, i.e. non-zero constant with respect to
$\rho$. The proof is similar to the one in~\cite[Appendix
  A]{Kumar:JIT09} and omitted here for brevity. We thus have
\begin{align}
P_{\text{out}}  &\,\dot{\geqslant}\;  \prob\bigg(\hat{\gamma} \leqslant 2^{\frac{R}{N}} -1\bigg) \nonumber\\
&= \prob\bigg(\hat{\gamma} \leqslant 2^{\frac{R}{N}} -1  \bigg| \mathcal{L}  \bigg) \prob(\mathcal{L}) \prob\bigg(\hat{\gamma} \leqslant 2^{\frac{R}{N}} -1  \bigg| \bar{\mathcal{L}}  \bigg) \prob \big(\bar{\mathcal{L}}\big) \nonumber\\
&\geqslant  \prob\bigg(\hat{\gamma} \leqslant 2^{\frac{R}{N}} -1  \bigg| \bar{\mathcal{L}}  \bigg) \prob(\bar{\mathcal{L}}) \nonumber\\
&\doteq  \prob\bigg(\hat{\gamma} \leqslant 2^{\frac{R}{N}} -1  \bigg| \bar{\mathcal{L}}  \bigg) \label{eq:LowerOutBoundWFZf} 
\end{align}
where~\eqref{eq:LowerOutBoundWFZf} holds since
$\prob(\bar{\mathcal{L}}) \doteq  O(1)$ as given
by~\eqref{eq:AsympValueAlpha2}.

We further bound the outage probability by bounding $\hat{\gamma}$ as
follows. Once again consider the two summations in the denominator
of~\eqref{eq:SINRTwoSum}. For the first summation
of~\eqref{eq:SINRTwoSum}, we have
\begin{align}
 &\sum_{j=1}^{N}\frac{\rho^{1-\alpha_j}}{(\rho^{1-\alpha_j}+N\,)^2} \doteq  \sum_{\alpha_j<1}\frac{1}{\rho^{1-\alpha_j}} +  \sum_{\alpha_j>1}\rho^{1-\alpha_j} \nonumber\\
&\doteq\begin{cases}
\rho^{-(1-\alpha_{\max})} &\text{$|\bar{\mathcal{L}}|=N$ } \\
\max(\rho^{-1+\alpha'},\rho^{1-\alpha''}) \;\dot{\geqslant}\; \rho^{1-\alpha_{\max}}  &\text{$1\leqslant |\bar{\mathcal{L}}| < N$}
\end{cases}\label{eq:SINRLowrBounSecSum}
\end{align}
where the bound in the second line~\eqref{eq:SINRLowrBounSecSum} is
true because
\[
\sum_{\alpha_j<1}\frac{1}{\rho^{1-\alpha_j}} +
\sum_{\alpha_j>1}\rho^{1-\alpha_j} \geqslant
\sum_{\alpha_j>1}\rho^{1-\alpha_j}\doteq \rho^{1-\alpha_{\max}}
\]

Using~\eqref{eq:EqAsympSINREig} and~\eqref{eq:SINRLowrBounSecSum} to
bound $\hat{\gamma}$
Substituting back in~\eqref{eq:SINRTwoSum} gives:
\begin{align}
\breve{\gamma}&\dot{\leqslant} \begin{cases}
  {\rho^{1-\alpha_{\max}}}\,\,=\rho\lambda_{\min} &\text{$|\bar{\mathcal{L}}|=N$
  } \\ {\rho^{1-\alpha_{\max}}}=\rho\lambda_{\min}  &\text{$1\leqslant
    |\bar{\mathcal{L}}| < N$ }
\end{cases}\nonumber\\
&\triangleq\breve{\breve{\gamma}} 
\label{eq:SINRWFPZFAsymptEnd2}
\end{align}

Thus the outage bound in~\eqref{eq:LowerOutBoundWFZf} can be then evaluated as we did for the upper bound
\begin{align}
P_{out} &\leqslant \prob\bigg(\hat{\gamma}\geqslant 2^{\frac{R}{N}} -1  \bigg| \bar{\mathcal{L}}  \bigg)\nonumber\\&\leqslant \prob\bigg(\breve{\breve{\gamma}}\leqslant 2^{\frac{R}{N}} -1  \bigg| \bar{\mathcal{L}}  \bigg) \nonumber\\
&< \prob\bigg(\breve{\breve{\gamma}} \leqslant 2^{\frac{R}{N}} -1  \bigg| |\mathcal{B}|=0   \bigg)\prob\big( |\mathcal{B}|=0 \big) +\nonumber\\
&\quad\quad\prob\bigg(\breve{\breve{\gamma}} \leqslant 2^{\frac{R}{N}} -1  \bigg| \bar{\mathcal{L}} , 0 <
    |\bar{\mathcal{B}}| < N  \bigg)\prob \big( |\bar{\mathcal{L}}| < N \big) \nonumber\\ 
&\doteq \prob \big( \lambda_{\min} \leqslant \rho^{-1}  \big)O(1)+\prob \big( \lambda_{\min} \leqslant \rho^{-1}  \big)O(1) \label{eq:BigO}\\
&\doteq \prob \big( \lambda_{\min} \leqslant \rho^{-1}  \big) \nonumber\\
&\doteq \rho^{-(M-N+1)}. \label{eq:OutBoundAlphaGreaOne3}
\end{align}
where~\eqref{eq:BigO} follows as a direct result of
Lemma~\ref{Lemma:AsympDist}. From~\eqref{eq:OutBoundAlphaGreaOne2}
and~\eqref{eq:OutBoundAlphaGreaOne3}, we conclude that the diversity
of MIMO system using TxWF precoder and ZF equalizer is
\begin{equation}
d^{WFP-ZF}=M-N+1. \nonumber
\end{equation}

\subsubsection{Regularized Zero Forcing Precoding}
\label{section:RZF-precoding}

The ZF equalizer is given by~\eqref{eq:ZFEqChWF} where the composite
channel $ \mathbb{H} = \bH \bT \nonumber$.
The received signal to noise ratio of
the $k$-th output symbol of the ZF filter as
\begin{align}
\gamma_{k} &= \frac{\rho \;\beta^2}{N\,R_{\tilde{n}}(k,k)}\nonumber\\
 &= \frac{\rho/N}{  \sum_{j=1}^{N}\frac{\lambda_j}{(\lambda_j+N\,)^2}      \sum_{l=1}^{N} \big(\frac{\lambda_l+N}{\lambda_l}\big)^2 |u_{kl}|^2 }\label{eq:SINRRZFPZFEq}.
\end{align}

The process of obtaining lower and upper bound has many similarities
with the developments of Section~\ref{section:WienerZF}, therefore we
omit many of the steps in the interest of brevity by referring to the
previous developments.


We begin with the outage upper bound, which is developed in a manner similar to~\eqref{eq:OutProbWFPZFEqu}.
\begin{align}
P_{\text{out}}&= \prob\bigg( \sum_{k=1}^{N}\log (1+\gamma_{k}) \leqslant R\bigg) \nonumber\\
&\leqslant  \prob\bigg( \sum_{k=1}^{N}\log (1+\hat{\gamma}) \leqslant R\bigg) \nonumber\\
&= \prob\bigg(\hat{\gamma} \leqslant 2^{\frac{R}{N}} -1\bigg) \label{eq:OutProbRZFTxZFEqu}
\end{align}

where
\begin{align}
\hat{\gamma} &= \frac{\rho/N}{  \sum_{j=1}^{N}\frac{\lambda_j}{(\lambda_j+N\,)^2}      \sum_{l=1}^{N} \big(\frac{\lambda_l+N}{\lambda_l}\big)^2  }\nonumber\\
&= \frac{\rho/N}{  \sum_{j=1}^{N}\frac{\rho^{-\alpha_j}}{(\rho^{-\alpha_j}+N\,)^2}      \sum_{l=1}^{N} \big(\frac{\rho^{-\alpha_l}+N}{\rho^{-\alpha_l}}\big)^2  }\nonumber\\
&\doteq \frac{\rho/N}{  \sum_{j=1}^{N}\rho^{-\alpha_j} \sum_{l=1}^{N} \rho^{2\alpha_l} } \\
&\,\dot{\geqslant}\; \frac{\rho/N}{  \sum_{l=1}^{N} \rho^{2\alpha_l} } \nonumber\\
&\doteq \frac{\rho/N}{  \rho^{2\alpha_{max}} }.
\label{eq:BoundSINRRZFPZFEq}
\end{align}

Thus the outage in~\eqref{eq:OutProbRZFTxZFEqu} can be bounded as 
\begin{align}
P_{\text{out}} &\leqslant \prob\bigg(\hat{\gamma} \leqslant 2^{\frac{R}{N}} -1\bigg) \nonumber\\
&\,\dot{\leqslant} \;\prob \bigg( \frac{\rho/N}{  \rho^{2\alpha_{max}} }\leqslant 2^{\frac{R}{N}} -1   \bigg) \nonumber\\
&\doteq \prob ( \lambda_{\min} \leqslant \rho^{-0.5}  ) \nonumber\\
&\doteq \rho^{-\frac{1}{2}(M-N+1)}. \label{eq:FracDiv}
\end{align}


We now turn to the lower bound, which is obtained in the same manner
as~\eqref{eq:LowerOutBoundWFZf}:
\begin{align}
P_{\text{out}}&= \prob\bigg( \sum_{k=1}^{N}\log (1+\gamma_{k}) \leqslant R\bigg) \nonumber\\
&\,\dot{\geqslant}\;  \prob\bigg( \sum_{k=1}^{N}\log (1+\breve{\gamma}) \leqslant R\bigg) \nonumber\\
&= \prob\bigg(\breve{\gamma} \leqslant 2^{\frac{R}{N}} -1\bigg) \label{eq:OutProbRZFTxZFEquLB}
\end{align}
where
\begin{align}
\breve{\gamma} &= \frac{\rho/N}{  \sum_{j=1}^{N}\frac{\lambda_j}{(\lambda_j+N\,)^2}      \sum_{l=1}^{N} \big(\frac{\lambda_l+N}{\lambda_l}\big)^2 \epsilon }\nonumber\\
&= \frac{\rho/N}{  \sum_{j=1}^{N}\frac{\rho^{-\alpha_j}}{(\rho^{-\alpha_j}+N\,)^2}      \sum_{l=1}^{N} \big(\frac{\rho^{-\alpha_l}+N}{\rho^{-\alpha_l}}\big)^2 \epsilon }\nonumber\\
&\doteq \frac{\rho/N}{  \sum_{j=1}^{N}\rho^{-\alpha_j} \sum_{l=1}^{N}  \epsilon \rho^{2\alpha_l}} \nonumber\\
&\leqslant \frac{\rho/N}{  \rho^{-\alpha_j} \sum_{l=1}^{N}  \epsilon \rho^{2\alpha_l}} 
\qquad \text{for arbitrary $j$}\nonumber\\
&\doteq \frac{\rho/N}{     \epsilon\; \rho^{-\alpha_{j}}  \rho^{2\alpha_{\max}} }\nonumber\\
&= \frac{\rho/N\lambda^2_{\min}}{ \epsilon\,\lambda_{j} } \nonumber\\
&\triangleq \breve{\breve{\gamma}}.
\label{eq:BoundSINRRZFPZFEqLB}
\end{align}

Let $C_1= (2^{\frac{R}{N}}-1)\,\epsilon \, N$, $C_2=C_1\xi$ where
$\xi$ is a fixed positive constant (independent of $\rho$), we have
\begin{align}
P_{\text{out}}&\;\dot{\geqslant}\;\prob\bigg({\breve{\gamma}} \leqslant 2^{\frac{R}{N}} -1\bigg) \nonumber\\
&\;\dot{\geqslant}\;\prob\bigg(\breve{\breve{\gamma}} \leqslant 2^{\frac{R}{N}} -1\bigg) \nonumber\\
&\;\dot{\geqslant}\; \prob \bigg( \frac{\rho\lambda^2_{\min}}{ \lambda_{j} } \leqslant C_1  \bigg)\nonumber\\
&\geqslant \prob \bigg( \frac{\rho\lambda^2_{\min}}{ \lambda_{j} } \leqslant C_1 \bigg| \lambda_{j} \geqslant \xi \bigg) \prob \big(\lambda_{j} \geqslant \xi \big) \nonumber\\
&\geqslant  \prob \bigg( \rho\lambda^2_{\min} \leqslant C_2 \bigg) \prob \big(\lambda_{j} \geqslant \xi \big) \nonumber\\
&\doteq  \prob \bigg( \rho\lambda^2_{\min} \leqslant C_2 \bigg). \label{eq:ExpIneqRZFZF}
\end{align}
The exponential inequality~\eqref{eq:ExpIneqRZFZF} holds because
$\prob \big(\lambda_{j} \geqslant \xi \big)=O(1)$, as proved in
Appendix~\ref{appendix:ExponentialInequality}. We thus conclude:
\begin{equation}
d^{RZFP-ZF}=\frac{1}{2}(M-N+1). \nonumber
\end{equation}

\begin{remark}
We note that the diversity of regularized zero-forcing precoder
together with a zero-forcing equalizer can be fractional. To our
knowledge this is the first instance of fractional diversity
uncovered in the literature.
\end{remark}

\subsubsection{Matched Filter Precoding}

In this case, the composite channel is
\begin{equation}
\mathbb{H} = \bH \bT =  \beta\bH \bH^H. \nonumber
\end{equation}

The noise correlation matrix is given by
\begin{align}
R_{\tilde{n}} = [\mathbb{H}^H\mathbb{H}]^{-1} 
=\frac{1}{ \beta^2}\;[(\bH\bH^H)^2]^{-1}
=\frac{1}{ \beta^2}\;(\bU\Lambda^2\bU^H)^{-1}. \nonumber
\end{align}

Thus
\begin{equation}
R_{\tilde{n}}(k,k) = \frac{1}{ \beta^2}\sum_{l=1}^{B} \frac{1}{\lambda^2_l} |u_{kl}|^2  \label{eq:GetNoiseCov2}
\end{equation}

The precoder normalization factor $\beta=1/\sqrt{\eta}$, where $\eta$ is given by 
\begin{equation}
\eta=\text{tr}\big[\bH\bH^H]=\sum_{l=1}^{N}\lambda_l \nonumber
\end{equation}

The signal to noise ratio of the $k$-th symbol of the ZF
filter is
\begin{align}
\gamma_{k} &= \frac{\rho}{N\,R_{\tilde{n}}(k,k)}\nonumber\\
 &= \frac{\rho/N}{  \sum_{j=1}^{N} \lambda_j     \sum_{l=1}^{N} \frac{1}{\lambda^2_l} |u_{kl}|^2 }
\label{eq:SINRMFPZFEq}.
\end{align}

Notice that the SINR $\gamma_{k}$ in~\eqref{eq:SINRMFPZFEq} is similar
to the SINR $\gamma_{k} $ of the RZF precoding system with ZF equalizer
given by~\eqref{eq:SINRRZFPZFEq}. The only difference is the term
$\lambda_k+N$ which, when applying the transformation of
$\lambda_k=\rho^{-\alpha_k}$, has no effect on the diversity analysis
as detailed in the previous section. We then conclude that the
diversity of the MIMO system applying MF precoder and ZF equalizer is
the same as the diversity of the RZF precoder with ZF equalizer. Thus:
\begin{equation}
d^{MFP-ZF}=\frac{1}{2}(M-N+1). \label{eq:DivMFZFeq}
\end{equation}

\subsection{MMSE equalizer}
\label{sec:MMSEEq}

The MMSE equalizer has better performance compared to ZF and is
therefore widely popular. We investigate the diversity of MIMO systems
that deploy different precoders at the transmitter and MMSE equalizer
at the receiver.

\subsubsection{MFTx Precoding}

The MFTx precoder, $\bP_{MFP}$ , is given by~\eqref{eq:MFPrecRes}. The MMSE equalizer for the precoded channel is given by
\begin{equation}
\bW_{MMSE}= \bigg[ \mathbb{H}^H\mathbb{H}+N\rho^{-1}\bI \bigg]^{-1} \mathbb{H}^H \label{eq:MMSEEqwithCompsiteCh}
\end{equation}
where $\mathbb{H}=\bH \,\bP_{MFP}=\beta_{MFP} \bH\bH^H $ and $\beta_{MFP}$ is given by~\eqref{eq:ScalarMFPrec}.

The SINR at the output of the MMSE filter is given by~\cite{Verdu98:book}
\begin{align}
\gamma_k &=\frac{\rho}{N} \bh_k \bigg[ \bI +\frac{\rho}{N} \mathbb{H}_k \mathbb{H}_k^H \bigg]^{-1} \bh_k\nonumber\\
&=\frac{1}{\bigg[  \bI +\frac{\rho}{N}  \mathbb{H}^H \mathbb{H}\bigg]^{-1}_{kk}} -1 \label{eq:SINRMMSEMFPrec}
\end{align}
where $\mathbb{H}_k$ is a submatrix of $\mathbb{H}$ obtained by
removing the $k$-th column, $\bh_k$.

The diversity analysis of the precoded system uses some results from
the un-precoded MMSE MIMO equalizers~\cite{Hesham:ISIT10}, which we quote in the following lemma.

\begin{lemma}
\label{lemma:ReviewMMSEResult}
consider a quasi-static Rayleigh fading MIMO channel $\bar{\bH}\in
\mathbb{C}^{M\times N}$ ($M \geqslant N$), the outage probability of
the MMSE receiver satisfies
\begin{align}
P_{out} &\doteq \prob \bigg( \text{tr} (\bI+\frac{\rho}{N}\bar{\bH}^H\bar{\bH})^{-1} \geqslant N2^{-\frac{R}{N}}  \bigg)\label{eq:ReviewMMSEEqout}\\
&= \prob \bigg( \sum_{k=1}^{N}\frac{1}{1+\frac{\rho}{N}\lambda'_k} \geqslant N2^{-\frac{R}{N}}  \bigg)\label{eq:ReviewMMSEEq}\\
&\doteq \rho^{-d^{MMSE}} \label{eq:MMSEReview}
\end{align}
where $\{\lambda'_k\}$ are the eigenvalues of $\bar{\bH}$ and
$d^{MMSE}$ is given by~\eqref{eq:DivMMSEAndWFP}.
 
\end{lemma}

Substituting $\lambda'_k=\rho^{-\alpha'_k}$, we have
\begin{equation}
 \frac{1}{1+\frac{\rho}{N}\lambda'_k} \doteq
 \begin{cases}
 \rho^{\alpha_k'-1} &\text{$\alpha_k' < 1$ } \\
 1 &\text{$\alpha_k' > 1$ }
\end{cases}\label{eq:AsympSINRRZFAprox4}
\end{equation}
thus the term $\frac{1}{1+\rho\lambda'_k/N}$ is either zero or one at
high SNR, and therefore to characterize the sum
in~\eqref{eq:ReviewMMSEEq} at high SNR we count the number of ones, or
equivalently the number of $\alpha'_k >1$. Hence the outage
probability reduces to~\cite{Hesham:ISIT10}

\begin{equation}
P_{out} \doteq \prob \bigg( \sum_{\alpha'_k>1}1 = \big\lceil N2^{-\frac{R}{N}} \big\rceil \bigg).  \label{eq:OutProbReviewMMSE}
\end{equation}

Now we apply the matched filter precoder.
Similarly to~\eqref{eq:ReviewMMSEEqout}, the outage portability  is given by
\begin{align}
P_{out} &\doteq \prob \bigg( \text{tr} (\bI+\frac{\rho}{N}\mathbb{H}\mathbb{H}^H)^{-1} \geqslant N2^{-\frac{R}{N}}  \bigg)\label{eq:MFPMMSEEqout1}\\
&= \prob \bigg( \sum_{k=1}^{N}\frac{1}{1+\frac{\rho}{N\eta}\lambda^2_k} \geqslant N2^{-\frac{R}{N}}  \bigg)\label{eq:MFPMMSEEqout2}
\end{align}
where we have used
$\mathbb{H}\mathbb{H}^H=\frac{1}{\eta}(\bH\bH^H)^2=\frac{1}{\eta}
\bU\Lambda^2\bU^H$ to obtain~\eqref{eq:MFPMMSEEqout2},and
$\{\lambda_k\}$ are the eigenvalues of the Wishart matrix $\bH\bH^H$. The scaling factor $\eta=\text{tr}(\bH\bH^H)=\sum_{l=1}^N \lambda_l$. 

We begin with a hypothetical precoder whose transmit power is not
normalized, i.e., $\eta=1$. The outage probability of this un-normalized
precoder is similar to that of the MMSE receiver with no precoding at
the transmitter, as given in~\eqref{eq:MMSEReview}, except that the
eigenvalues are now squared. Thus similarly
to~\eqref{eq:AsympSINRRZFAprox4}, we have the exponential inequality
\begin{equation}
 \frac{1}{1+\frac{\rho}{N}\lambda^2_k} \doteq
 \begin{cases}
 \rho^{2\alpha_k-1} &\text{$\alpha_k < 0.5$ } \\
 1 &\text{$\alpha_k > 0.5$ }
\end{cases}\label{eq:AsympSINRRZFAprox5}.
\end{equation}

The analysis of~\cite{Hesham:ISIT10} then follows and we have
\begin{equation}
d= \frac{1}{2} \bigg(\lceil N2^{-\frac{R}{N}} \rceil^{2}+ (M-N)\lceil M
2^{-\frac{R}{N}}\rceil\bigg) \label{eq:DivMMSEAndMFPUnc}.
\end{equation}

We conclude that the un-normalized matched filter precoding with MMSE
receiver results in $50\%$ diversity loss compared to MMSE receiver
with no transmit precoding. 


For the normalized precoder, we begin with the outage probability
in~\eqref{eq:MFPMMSEEqout2}. Assume $\alpha_1 \geqslant \alpha_2
\cdots \geqslant \alpha_N$, the sum term in~\eqref{eq:MFPMMSEEqout2}
is given by
\begin{align}
\sum_{k=1}^{N}\frac{1}{1+\frac{\rho}{N\eta}\lambda^2_k} &= \sum_{k=1}^{N}\frac{\eta}{\eta+\frac{\rho}{N}\lambda^2_k}\nonumber\\
 &=\sum_{k=1}^{N}\frac{\sum_l\rho^{-\alpha_l}}{\sum_l\rho^{-\alpha_l}+\frac{\rho}{N}\rho^{-2\alpha_k}} \nonumber\\
&\doteq  \sum_{k=1}^{N}\frac{\rho^{-\alpha_{N}}}{\rho^{-\alpha_N}+\rho^{1-2\alpha_k}} \label{eq:MFPMMSEEqSumTerm}.
\end{align}
where we have used the fact that the $\sum_l \rho^{-\alpha_k}$ is
dominated by the maximum element at high SNR. It is easy to see that
the terms of~\eqref{eq:MFPMMSEEqSumTerm} are either one or zero at high SNR, depending on
whether $\rho^{-\alpha_{N}}$ asymptotically dominates
$\rho^{1-2\alpha_k}$ or vice versa. These two cases are delineated with the threshold
$\alpha_k \lessgtr 0.5 \max (1\,,\,\alpha_N + 1) $, or, considering that $\alpha_N$ is positive, $\alpha_k \lessgtr 0.5 (\alpha_N + 1) $.
Thus at high SNR, the outage probability is evaluated by counting the
ones
\begin{align}
P_{out} &\doteq \prob \bigg(
\sum_{k=1}^{N}\frac{1}{1+\frac{\rho}{N\eta}\lambda^2_k} \geqslant
N2^{-\frac{R}{N}} \bigg) \nonumber \\
&\doteq \prob \bigg( \sum_{\alpha_k> 0.5\,(\alpha_N + 1)}1 \geqslant
N2^{-\frac{R}{N}}  \bigg) \nonumber\\
&\doteq \prob \bigg( \sum_{\alpha_k > 0.5\,(\alpha_N + 1) }1 = L \bigg)
\label{eq:MFPMMSEEqoutNew}
\end{align}
where $L=\big\lceil N2^{-\frac{R}{N}} \big\rceil$. The conversion from
inequality to equality in equation~\eqref{eq:MFPMMSEEqoutNew} follows
from arguments developed in~\cite[Section III-A]{Hesham:ISIT10} .

Therefore, the outage probability is asymptotically evaluated by:
\begin{align}
P_{out} &\doteq \int_{{\mathcal S}^+} \prob (\balpha)  \; d\balpha \label{eq:MMSEMFPrecCountOnes2}
\end{align}
where $ \prob (\balpha)$ is the joint distribution of the ordered
$\alpha_1 \geqslant \cdots \geqslant \alpha_N$ and the region of integration is defined as ${\mathcal S}^+ = {\mathcal S} \cap {\mathbb R}^{N+}$, where $\mathcal S$ is given as follows:
\begin{itemize}
\item If $L=N$, then we seek the probability that $\alpha_k >
  \frac{1}{2}(\alpha_N+1)$ for $k=1,\ldots,N$, which implies $\alpha_N
  \in (1,\infty)$. Thus the integration region can be tightly
  represented as:
\[
{\mathcal S} = \big\{ \alpha_N>1 \; ,\; \min_{1\le k < N} \alpha_k > 0.5(\alpha_N+1) \big\}
\]

\item If $L<N$, then we seek the joint probability that $\alpha_k > \frac{1}{2}(\alpha_N+1)$ for $k=1,\ldots, L$ and $\alpha_k \le \frac{1}{2}(\alpha_N+1)$ for $k=L+1,\ldots, N$, implying $\alpha_N \in (0,1)$. Thus the region of integration is represented as:
\[
{\mathcal S} = \big\{ \alpha_N< 1 \; ,\; \min_{1<k\le L} \alpha_k > 0.5(\alpha_N+1) \; ,\; \max_{L<k<N}\alpha_k <0.5(\alpha_N+1) \big\}
\]
\end{itemize}

Using methods similar to~\cite{Zheng:JIT03} and~\cite[Eq (18) -
  (20)]{Hesham:ISIT10}, exponential equality relations can be used to
reduce the integrand to the following:
\begin{align}
P&_{out} \doteq \int_{{\mathcal S}^+} \ \prod_{k}\rho^{-(2k-1+M-N)\alpha_k} \; d(\balpha) \label{eq:MMSEMFPrecCountOnes3}
\end{align}
First we consider $L=N$. The probability expression is evaluated by simply taking the integral over all variables except $\alpha_N$, and then taking an integral over $\alpha_N$.
\begin{align}
P&_{out} \doteq  \int_{\alpha_N=1}^\infty \rho^{-(2N-1+M-N)\alpha_N} \nonumber\\
 &\qquad\times \prod_{k=1}^{N-1}\rho^{-(2k-1+M-N)(0.5+0.5\alpha_N)} d(\balpha)  \label{eq:MMSEMFPrecCountOnes8}\\
 &\doteq \prod_{k=1}^{N}\rho^{-(2k-1+M-N)} \nonumber\\
&= \rho^{\sum_{k=1}^{N}-(2k-1+M-N)} \label{eq:MMSEMFPrecCountOnes9}\\
&= \rho^{-MN}. \label{eq:MMSEMFPrecCountOnes10}
\end{align}
When $L < N$, we repeat the same integration strategy.
\begin{align}
P_{out} &\doteq \int_{\alpha_N=0}^{1} \rho^{-(2N-1+M-N)\alpha_N} \nonumber\\
&\qquad\times \prod_{l=L+1}^{N} \bigg(1-\rho^{-(2l-1+M-N)(0.5+0.5\alpha_N)}\bigg) 
 \nonumber\\
 &\qquad\times \prod_{k=1}^{L}\rho^{-(2k-1+M-N)(0.5+0.5\alpha_N)} d(\balpha) \label{eq:MMSEMFPrecCountOnes4}\\
&\doteq \int_{\alpha_N=0}^{1} \rho^{-(2N-1+M-N)\alpha_N} \nonumber\\
 &\qquad\times \prod_{k=1}^{L}\rho^{-(2k-1+M-N)(0.5+0.5\alpha_N)} d(\balpha) \label{eq:MMSEMFPrecCountOnes5}\\
 &\doteq \prod_{k=1}^{L}\rho^{-\frac{1}{2}(2k-1+M-N)} \nonumber\\
&= \rho^{\sum_{k=1}^{L}-\frac{1}{2}(2k-1+M-N)} \nonumber\\
&= \rho^{-\frac{1}{2}(L^2+(M-N)L)} \label{eq:MMSEMFPrecCountOnes6}
\end{align}
In deriving~\eqref{eq:MMSEMFPrecCountOnes4}
and~\eqref{eq:MMSEMFPrecCountOnes5} we have used $\int_{a}^{b}
\rho^{-c_k\alpha_k} d(\alpha_k) \doteq \rho^{-ac_k}
$~\cite{Hesham:ISIT10}. Equations~\eqref{eq:MMSEMFPrecCountOnes10}
and~\eqref{eq:MMSEMFPrecCountOnes6} show that the system exhibits two
distinct diversity behaviors based on whether $L =\lceil N
2^{-\frac{R}{N}}\rceil < N$. We can solve to find the boundary of the
two regions $R=N \log \frac{N}{N-1}$. To summarize:
\begin{align}
&d^{MFP-MMSE} =\nonumber\\
& \begin{cases}
 \frac{1}{2}\big(\lceil N2^{-\frac{R}{N}} \rceil^{2}+ (M-N)\lceil M
2^{-\frac{R}{N}}\rceil\big) &R > N \log \frac{N}{N-1}\\
MN &\text{otherwise}  
\end{cases}\label{eq:DivMMSEMatchedPrecRes}.
\end{align}

\begin{remark}
The outcome is interesting for its practical implications: An MMSE
receiver working with matched-filter precoding will suffer a
significant diversity loss compared to an MMSE receiver without
precoding, except for very low rates corresponding to $R<N
\log\frac{N}{N-1}$, where the combination of MMSE receiver with
matched filter precoding has exactly the same diversity as the MMSE
receiver alone.
\end{remark}

\begin{remark}
Recall that $R=N \log\frac{N}{N-1}$ is exactly the same threshold
below which matched filter precoding (without receiver-side
equalization) achieves full diversity.
\end{remark}

\subsubsection{WFTx Precoding}
\label{sec:WFPrecMMSEEq}

Using the Wiener filter precoding at the receiver results in the composite channel
\begin{equation}
\mathbb{H}= \bH\bP = \beta\bH\bH^H(\bH\bH^H+\rho^{-1}N\bI)^{-1}.\nonumber
\end{equation}
Using the eigen decomposition $\bH\bH^H=\bU\Lambda\bU^H$, it can be shown that
\begin{align}
\mathbb{H}^H\mathbb{H}= \beta^2\bU (\Lambda+\rho^{-1}N\bI)^{-2}\Lambda^2  \bU^H \label{eq:CompositeChWFPMMSEEq}
\end{align}

Similar to the case of MF precoder with MMSE receiver, the outage probability of WF precoder with MMSE receiver is given by (c.f.~\eqref{eq:MFPMMSEEqout1})
\begin{align}
P_{out} &\doteq \prob \bigg( \text{tr} (\bI+\frac{\rho}{N}\mathbb{H}\mathbb{H}^H)^{-1} \geqslant N2^{-\frac{R}{N}}  \bigg)\nonumber\\
&= \prob \bigg( \sum_{k=1}^{N}\frac{1}{1+\frac{\rho}{N\eta}\hat{\lambda}_k} \geqslant N2^{-\frac{R}{N}}  \bigg)\label{eq:WFPMMSEEqout2}
\end{align}
where $\{\hat{\lambda}_k\}$ are the eigenvalues of
$\mathbb{H}^H\mathbb{H}$ and $\eta$ is the scale
factor. Using~\eqref{eq:CompositeChWFPMMSEEq}, $\{\hat{\lambda_k}\}$
are given by
\begin{equation}
\hat{\lambda}_k=\frac{\lambda_k^2}{(\lambda_k+\rho^{-1}N)^2}, \quad k=1,\dots,N\label{eq:lambdaCompsitCh}
\end{equation}

The scale factor $\eta$ is calculated as in~\eqref{eq:NormFactor}
\[
\eta =\sum_{l=1}^{N}\frac{\lambda_l}{(\lambda_l+\rho^{-1}N\,)^2}.
\]
Thus the outage probability can be written as 
\begin{align}
P_{out} \doteq \prob \bigg( \sum_{k=1}^{N} \gamma_k \geqslant N2^{-\frac{R}{N}}  \bigg)\label{eq:WFPMMSEEqout3}
\end{align}
where
\begin{align}
\gamma_k &\triangleq \frac{1}{1+\frac{\rho}{N\eta}\hat{\lambda}_k} = \frac{\rho^{-1}\eta}{\rho^{-1}\eta+\frac{1}{N}\hat{\lambda}_k} = \frac{\rho^{-1}\eta}{\rho^{-1}\eta+\upsilon_k} \nonumber
\end{align}
where we define $\upsilon_k = \frac{1}{N}\hat{\lambda}_k$. We now proceed to express both $\rho^{-1}\eta$ and $\upsilon_k$ in terms of $\{\alpha_k\}$, the exponential orders of $\{\lambda_k\}$.
\begin{align}
\rho^{-1}\eta&=\sum_{l=1}^{N}\frac{\rho^{-1}\lambda_l}{(\rho^{-1}\lambda_l+N\,)^2}
=\sum_{l=1}^{N}\frac{\rho^{1-\alpha_l}}{(\rho^{1-\alpha_l}+N\,)^2} \nonumber\\ 
&\doteq \sum_{\alpha_l>1} \rho^{1-\alpha_l} + \sum_{\alpha_l<1} \rho^{\alpha_l-1} \label{eq:etaVal}
\end{align}
observe that all the terms in~\eqref{eq:etaVal} have negative exponent.
Using~\eqref{eq:lambdaCompsitCh}, 
\begin{align}
\upsilon_k&=\frac{1}{N}\frac{\rho^{-2\alpha_k}}{(\rho^{-\alpha_k}+\rho^{-1}N)^2} \nonumber\\
&= \frac{1}{N}\frac{\rho^{2(1-\alpha_k)}}{(\rho^{1-\alpha_k}+N)^2}\nonumber \\
& \doteq
 \begin{cases}
1 &\text{$\alpha_k < 1$ } \\
 \rho^{2(1-\alpha_k)} &\text{$\alpha_k > 1$ }
\end{cases}\label{eq:UpsilonVariable}.
\end{align}

From~\eqref{eq:etaVal} and~\eqref{eq:UpsilonVariable}, we see that
when $\alpha_k <1$ then $\upsilon_k + \rho^{-1}\eta \doteq \upsilon_k
\doteq 1 $. On the other hand, when $\alpha_k >1$ then 
\begin{align}
\upsilon_k + \rho^{-1}\eta &\doteq  \rho^{2(1-\alpha_k)}  + \sum_{\alpha_l>1} \rho^{1-\alpha_l} + \sum_{\alpha_l<1} \rho^{\alpha_l-1} \nonumber\\
&= \rho^{2(1-\alpha_k)}  +  \rho^{1-\alpha_k} + \sum_{\substack{\alpha_l>1 \\ l\ne k}} \rho^{1-\alpha_l} + \sum_{\substack{\alpha_l<1 \\l\ne k}} \rho^{\alpha_l-1} \nonumber\\
&\doteq \rho^{1-\alpha_k} + \sum_{\substack{\alpha_l>1 \\ l\ne k}} \rho^{1-\alpha_l} + \sum_{\substack{\alpha_l<1 \\l\ne k}} \rho^{\alpha_l-1} \label{eq:AproxSumRhoWFPMMSE}\\
&\doteq \rho^{-1}\eta
\end{align}
where~\eqref{eq:AproxSumRhoWFPMMSE} follows because $\alpha_k > 1$.
Thus we have
\begin{equation}
\gamma_k =\frac{\rho^{-1}\eta}{\rho^{-1}\eta+\upsilon_k} \doteq
 \begin{cases}
\rho^{-1}\eta&\text{$\alpha_k < 1$ } \\
1 &\text{$\alpha_k > 1$ }
\end{cases}\label{eq:GammaVariable}
\end{equation}
and $\rho^{-1}\eta$ has negative exponent thus vanishes at high SNR.

Observe that~\eqref{eq:GammaVariable} is similar
to~\eqref{eq:AsympSINRRZFAprox4} which corresponds to the case of the
MMSE-only system (i.e. with no precoding). Thus
substituting~\eqref{eq:GammaVariable} in the outage probability~\eqref{eq:WFPMMSEEqout3} and repeating the same analysis of the
MMSE-only system as in~\cite{Hesham:ISIT10}, we conclude that the
diversity of the MMSE receiver when using WFTx precoding is the same
as the diversity of the MMSE receiver with no linear precoding, which is given by~\eqref{eq:DivMMSEAndWFP}.

\subsubsection{RZF Precoding}
\label{sec:RZFPrecMMSEEq}

Using the Regularized Zero Forcing precoding at the receiver results
in the composite channel
\begin{equation}
\mathbb{H}= \bH\bP = \beta\bH\bH^H(\bH\bH^H+c\,\bI)^{-1}.\nonumber
\end{equation}
where $c$ is a fixed constant, $\beta=1/\eta$ and $\eta$ is
given by~\eqref{eq:NormFactor}
\begin{equation}
\eta =\sum_{l=1}^{N}\frac{\lambda_l}{(\lambda_l+c\,)^2} = \sum_{l=1}^{N}\frac{\rho^{-\alpha_l}}{(\rho^{-\alpha_l}+c\,)^2}. \label{eq:RZFPMMSEEqEta}
\end{equation}

Similar to~\eqref{eq:WFPMMSEEqout2}, the outage probability of RZF
precoder with MMSE receiver is given by
\begin{align}
P_{out} \doteq \prob \bigg( \sum_{k=1}^{N} \gamma_k \geqslant N2^{-\frac{R}{N}}  \bigg)\nonumber
\end{align}
and
\begin{align}
\gamma_k &\triangleq \frac{\eta}{\eta+\frac{\rho}{N}\bar{\lambda}_k} \nonumber
\end{align}
where $\{\bar{\lambda}_k\}$ are the eigenvalues of
$\mathbb{H}^H\mathbb{H}$ given by
\begin{equation}
\bar{\lambda}_k=\frac{\lambda_k^2}{(\lambda_k+c)^2}=\frac{\rho^{-2\alpha_k}}{(\rho^{-\alpha_k}+c)^2}, \quad k=1,\dots,N \label{eq:RZFPMMSEEqEigen}
\end{equation}

Notice that at high SNR we have
\begin{align}
\eta &\doteq  \sum_{l=1}^{N}\frac{\rho^{-\alpha_l}}{c^2}  \nonumber\\
\bar{\lambda}_k &\doteq \frac{\rho^{-2\alpha_k}}{c^2}. \nonumber
\end{align}

Thus the SINR is given by (c.f.~\eqref{eq:MFPMMSEEqSumTerm})
\begin{align}
\gamma_k \doteq \frac{ \sum_{l=1}^{N}\rho^{-\alpha_l}}{ \sum_{l=1}^{N}\rho^{-\alpha_l}+\rho^{-2\alpha_k}} &\doteq \frac{\rho^{-\alpha_{N}}}{\rho^{-\alpha_N}+\rho^{1-2\alpha_k}},\nonumber\\ &\quad k=1,\dots,N \nonumber
\end{align}
which are the same terms as in~\eqref{eq:MFPMMSEEqSumTerm}, implying that the outage probability of the MMSE receiver working with the regularized zero-forcing precoder is asymptotically the same as the outage probability of the MMSE receiver working with the matched filter precoder. This means:
\begin{align}
&d^{RZFP-MMSE} = d^{MFP-MMSE}.\nonumber\\
&= \begin{cases}
 \frac{1}{2}\big(\lceil N2^{-\frac{R}{N}} \rceil^{2}+ (M-N)\lceil M
2^{-\frac{R}{N}}\rceil\big) &R > N \log \frac{N}{N-1}\\
MN &\text{otherwise}  
\end{cases}
\end{align}

\section{Simulation Results}
\label{sec:SimRes}

\begin{figure}
\centering
\includegraphics[width=3.60in]{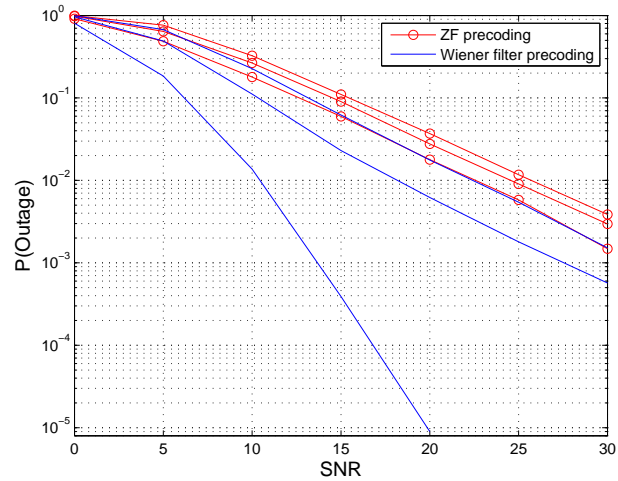}
\caption{Outage probability of the ZF and Wiener filtering precoded
  MIMO $2\times 2$ system for rates (left to right): $R=1.9, 2.5,$ and $3$
  b/s/Hz.}
\label{fig:Fig1}
\end{figure}
\begin{figure}
\centering
\includegraphics[width=3.60in]{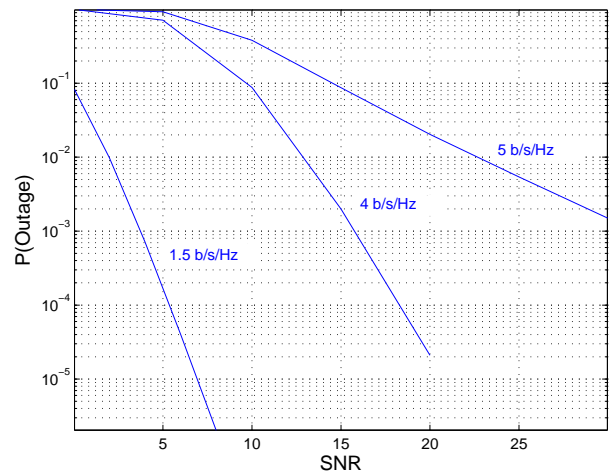}
\caption{Wiener precoded $3\times 3$ MIMO system.}
\label{fig:Fig2}
\end{figure}
\begin{figure}
\centering
\includegraphics[width=3.60in]{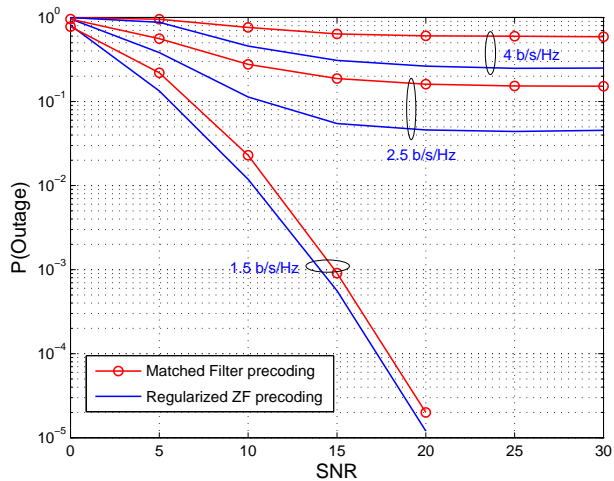}
\caption{MF and regularized ZF precoded
  $2\times 2$ MIMO system for rates (left to right): $R=1.9, 2.5,$
  and $4$ b/s/Hz.}
\label{fig:Fig3}
\end{figure}
\begin{figure}
\centering
\includegraphics[width=3.60in]{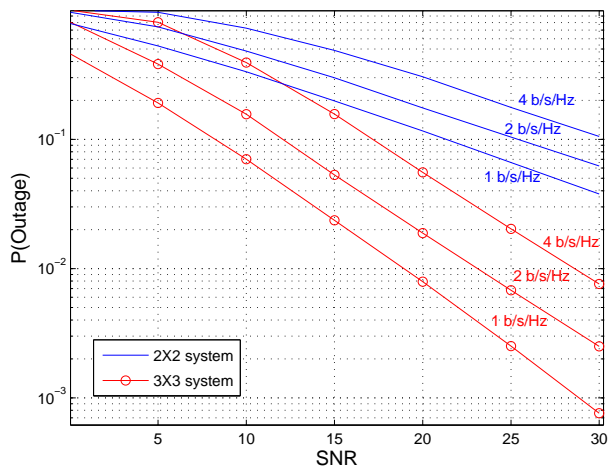}
\caption{MIMO system with matched filtering
  precoding and ZF equalization for rates (left to right): $R=1, 2,$
  and $4$ b/s/Hz.}
\label{fig:Fig4}
\end{figure}
\begin{figure}
\centering
\includegraphics[width=3.60in]{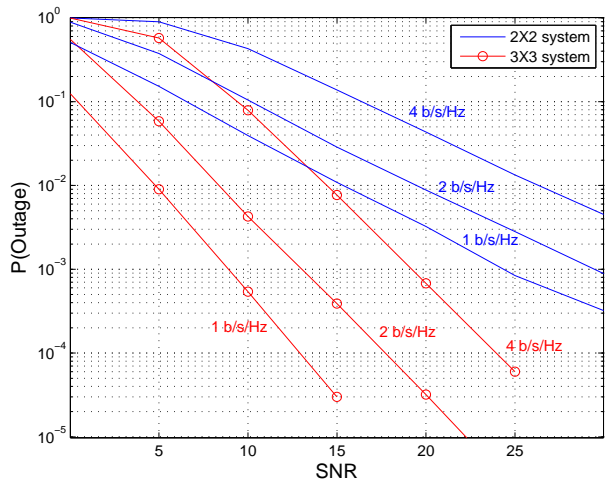}
\caption{Outage probability of MIMO system with Wiener filtering
  precoding and ZF equalization for rates (left to right): $R=1, 2,$
  and $4$ b/s/Hz.}
\label{fig:Fig5}
\end{figure}
\begin{figure}
\centering
\includegraphics[width=3.60in]{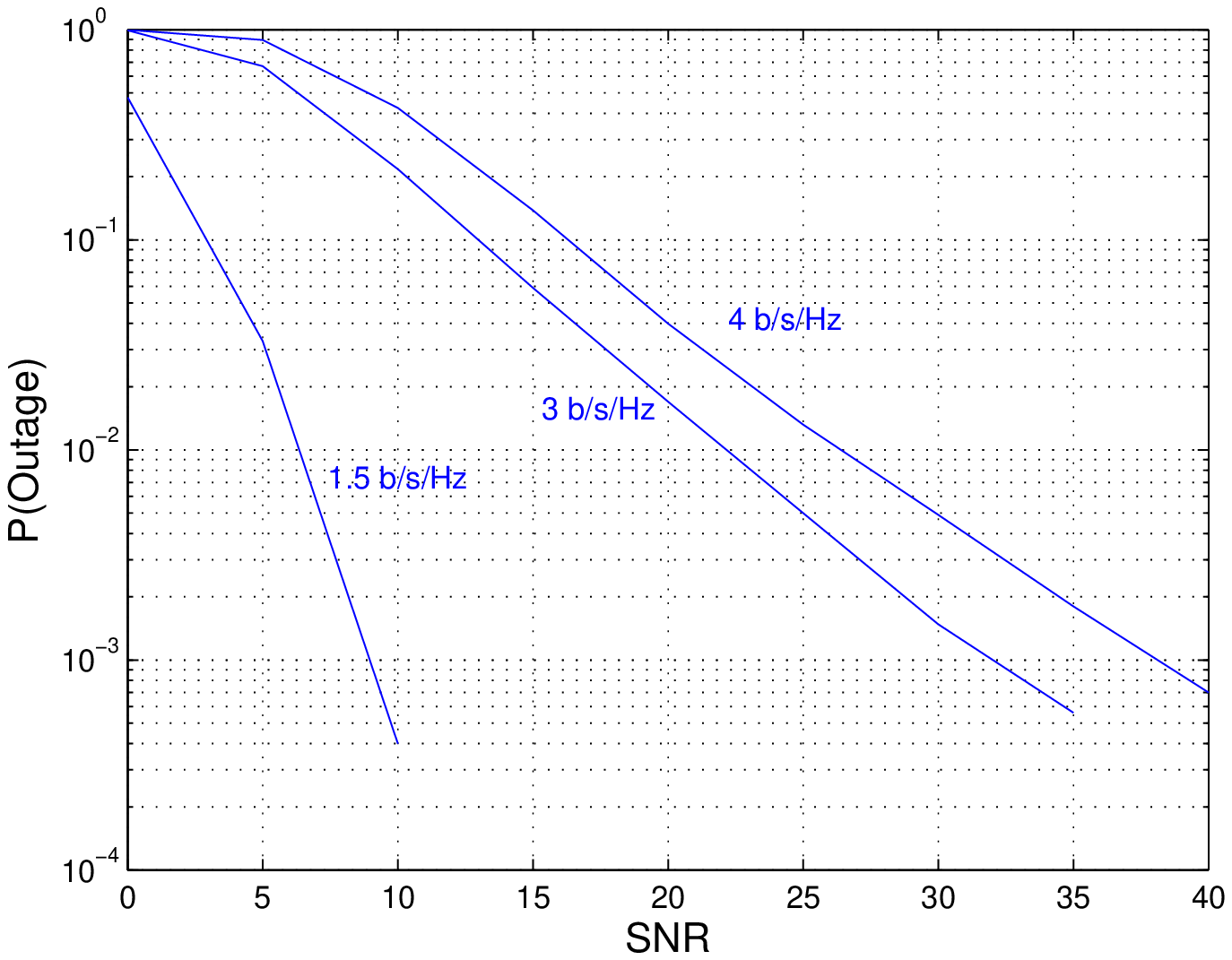}
\caption{2X2 MIMO system with Wiener filtering
  precoding and MMSE equalization for rates (left to right): $R=1.5, 3,$
  and $4$ b/s/Hz.}
\label{fig:Fig6}
\end{figure}
\begin{figure}
\centering
\includegraphics[width=3.60in]{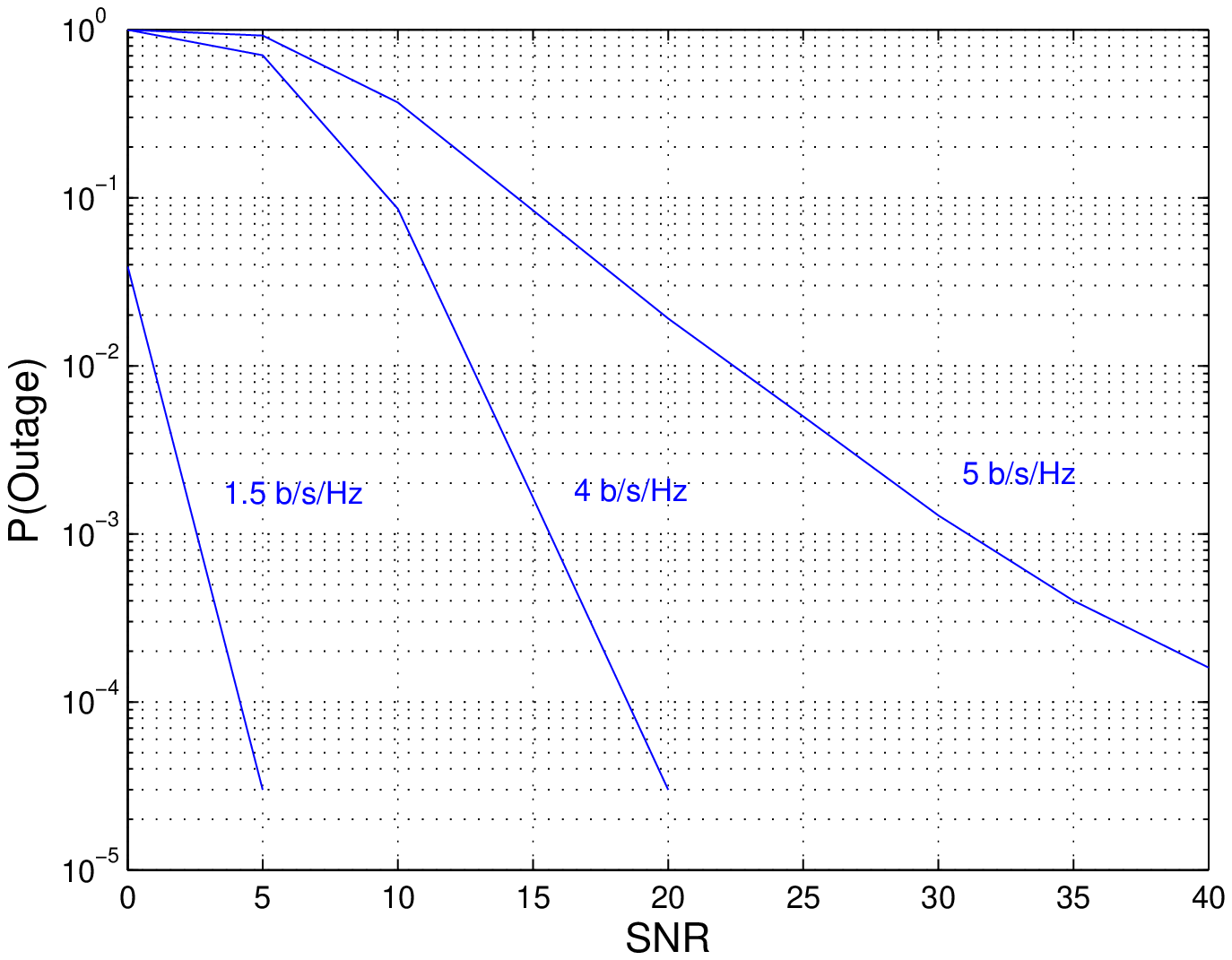}
\caption{3X3 MIMO system with Wiener filtering
  precoding and MMSE equalization for rates (left to right): $R=1.5, 4,$
  and $5$ b/s/Hz.}
\label{fig:Fig7}
\end{figure}
\begin{figure}
\centering
\includegraphics[width=3.60in]{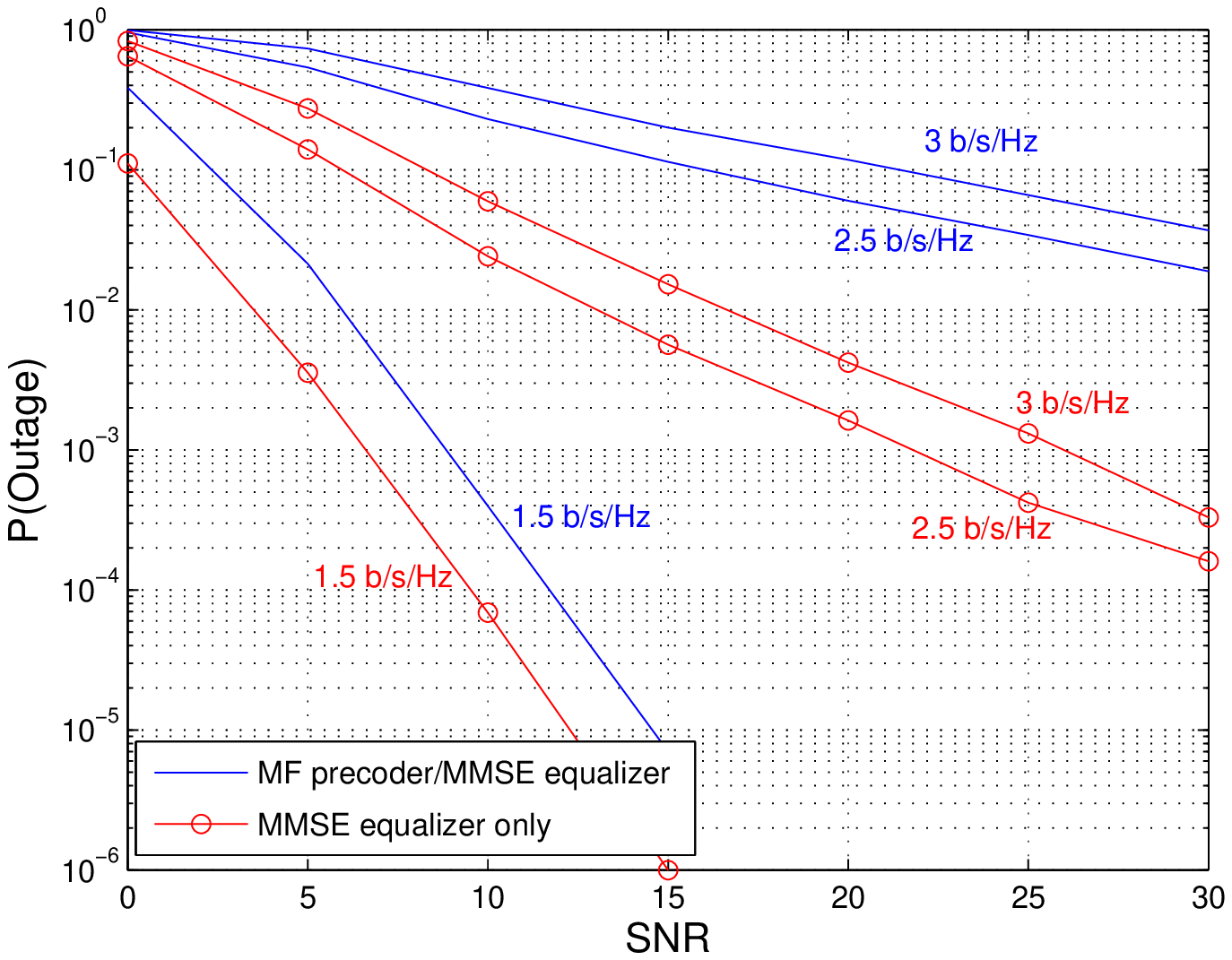}
\caption{2X2 MIMO system with MF precoding and
  MMSE equalization system  for rates (left to right): $R=1.5, 2.5,$
  and $3$ b/s/Hz.}
\label{fig:Fig8}
\end{figure}

This section produces numerical results for the outage probabilities
of ZF, regularized ZF (RZF), matched filter (MF) and Wiener precoding
systems. Figure~\ref{fig:Fig1} shows the outage probabilities of the
ZF and Wiener-filter precoded $2\times 2$ MIMO systems. The diversity
in the case of the ZF case is the same as the one predicted by the
DMT. In the case of Wiener precoding, the diversity is the same as the
one predicted by the DMT for high rate ($R$) values and it departs
from the DMT for low rate values. A complete diversity
characterization is given by~\eqref{eq:DivMMSEAndWFP} which is similar
to that of the MMSE MIMO
equalizer~\cite{Hesham:ISIT10}. Figure~\ref{fig:Fig2} shows outage
probabilities for a $3\times 3$ MIMO system with Wiener precoding. The
diversity for the rates $R=1.5, 4,$ and $5$ b/s/Hz is $9, 4$ and $1$
respectively. Figure~\ref{fig:Fig3} shows an error floor for the
regularized ZF and matched filtering precoded $2\times 2$ system at
high rates. However we observe that the maximum diversity is achieved
for any rate $R < 2 $
(c.f. Equation~\eqref{eq:rateErroFloor}). Figure~\ref{fig:Fig4} shows
outage probabilities for a $2\times 2$ and a $3\times 3$ MIMO system
with matched filter precoding and ZF equalization. The observed
diversity values are consistent with Eq.~\eqref{eq:DivMFZFeq}.
Figure~\ref{fig:Fig5} shows outage probabilities for a $2\times 2$ and
a $3\times 3$ MIMO system with Wiener filter precoding and ZF
equalization.  Figure~\ref{fig:Fig6} and Figure~\ref{fig:Fig7} show
outage probabilities for a $2\times 2$ and a $3\times 3$ MIMO system,
respectively, with Wiener filter precoding and MMSE equalization. The
diversity for the $3\times 3$ system is the same as the diversity of
the Wiener filtering precoding-only (c.f.  Figure~\ref{fig:Fig2}).

Figure~\ref{fig:Fig8} shows the outage probability of a $2\times 2$
MIMO system with matched filter precoding and MMSE equalization, which is consistent with Eq.~\eqref{eq:DivMMSEMatchedPrecRes}. We also plot the outage
probability of the MMSE MIMO equalizer (without any precoding) for
comparison.

\section{Conclusion}
\label{sec:Conclusion}

Linear precoders provide a simple and efficient processing, and have
been shown to be optimal in some scenarios~\cite{Caire:TIF99,
  Skoglund:SAC03, Taesang:SAC06}. This paper studies the high-SNR
performance of linear precoders.  It is shown that the zero-forcing
precoder under two common design approaches, maximizing the throughput
and minimizing the transmit power, achieves the same DMT as that of
MIMO systems with ZF equalizer. When a regularized ZF (RZF) precoder
(for a fixed regularization term that is independent of the
signal-to-noise ratio) or matched filter (MF) precoder is used, we
have $d(r)=0$ for all $r$, implying an error floor under all
conditions. It is also shown that in the fixed rate regime RZF and MF
precoding achieve full diversity up to a certain spectral efficiency,
while at higher spectral efficiencies they produce an error floor. If
the regularization parameter in the RZF is optimized in the MMSE
sense, the RZF precoded MIMO system exhibits a complex rate-dependent
behavior. In particular, the diversity of this system (also known as
Wiener filter precoding) is characterized by $d(R) = \lceil
N2^{-\frac{R}{N}} \rceil^{2}+ (M-N)\lceil N 2^{-\frac{R}{N}}\rceil$
where $M$ and $N$ are the number of transmit and receive
antennas. This is the same behavior observed in linear MMSE MIMO
receivers~\cite{Hesham:ISIT10}. Various results for the diversity in
the presence of {\em both} precoding and equalization have also been
obtained.


\appendix

\subsection{Pairwise error Probability (PEP) Analysis}
\label{appendix:PEP}
In this section we perform PEP analysis for the the zero-forcing (ZF)
and the regularized ZF (RZF) precoding systems. The presented analysis can be
easily extended to all other precoding systems.  The basic strategy is
to show the SNR exponent of outage probability bounds the SNR exponent
of PEP from both sides The PEP analysis follows
from~\cite{Tajer:WCOM10,Hesham:ISIT10}, with careful attention to the
system model given by Equation~\eqref{eq:SysMod}.

The lower bound immediately follows from~\cite[Lemma 3]{Tajer:WCOM10}
by recognizing that although it was developed for SISO block
equalization, nowhere in its development does it depend on the number
of receive antennas, therefore we can directly use it for our
purposes:
\begin{equation}
P_{err} \;\dot{\geqslant}\; P_{out}. \label{eq:LBOnPEP}
\end{equation}

The upper bound on PEP for the ZF/RZF precoding systems receiver is
developed using the union bound. Denote the channel outage event by
$O$ and the error event by $E$. The PEP is given by
\begin{align}
P_{\text{err}}&= P(E|O)\;P_{out} + P(E,\bar{O}) \nonumber \\
&\leqslant P_{out} + P(E,\bar{O}) .\label{eq:PEPUB}
\end{align}
In order to show that $P_{out}$ dominates the right hand side of
\eqref{eq:PEPUB}, it is
shown in~\cite{Hesham:ISIT10} that the probability
$P(E,\bar{O})$ can be bounded as  follows using the union bound
\begin{align}
\prob(E,\bar{O}) \quad &\dot{\leqslant}\quad 2^{Rl} e^{-\frac{\rho/N}{\sigma^2_{\tilde{\bn}}(k)}} \;\dot{\leqslant}\;\; \rho^{-MN}\label{eq:PEPAsym}
\end{align}
where $l$ is the codeword length and ${\sigma^2_{\tilde{\bn}}(k)}$ is
the variance of the interference plus noise signal $\tilde\bn$ in the
$k$-th receive stream~\footnote{~\cite{Tajer:WCOM10} analyzes linear
  receivers so $\tilde\bn$ is the $k$-th output filtered interference
  plus noise signals. By symmetry assumption all the equalizer outputs
  have equal noise variance.}.  The proof of~\cite{Tajer:WCOM10} does
not depend on the codeword length for both upper and lower PEP
bounds. The bound are tight and were confirmed by simulations for
outage and error probabilities.

We now show that a similar proof holds for regularized zero-forcing
(RZFP).  Recall that the outage probability of the RZFP can be upper
bounded by~\eqref{eq:BoundRZFdiv2}  
\begin{equation}
P_{out} \leqslant  \prob\big(  \frac{\nu}{\mu_{\max}}  \, \leqslant \,\Theta  \big)
\triangleq \;P_{out}^{b} \label{eq:UpperBoundSum3}
\end{equation}
  We will use~$P_{out}^{b}$ to further
  bound~\eqref{eq:PEPUB}. Moreover $ P(E,\bar{O})$ can be upper
  bounded by bounding the noise variance ${\sigma^2_{\tilde{\bn}}(k)}$ in~\eqref{eq:PEPAsym}
\begin{align}
\sigma^2_{\tilde{\bn}}(k) = P_I + P_n < P_T + 1\label{eq:sigmaBar}
\end{align}
where we have used the noise power $P_n=1$, and bound the interference
power by the total received power $P_T$. We will first consider the
case of RZF precoding since the case of ZF precoding can be easily
deduced from RZF by substituting setting the regularization parameter
$c=0$. For the RZF precoding system we use the $P_T$ given
by~\eqref{eq:TotSigPower} which can be simplified in a way similar to
earlier sections
\begin{align}
P_T &=\frac{\beta^2 \rho}{N}\sum_{l=1}^{N} \frac{\lambda_l^2}{(\lambda_l+c\,)^2} \nonumber\\ &= \frac{1}{\sum_{l=1}^{N}\frac{\lambda_l}{(\lambda_l+c\,)^2}}\frac{ \rho}{N}\sum_{l=1}^{N} \frac{\lambda_l^2}{(\lambda_l+c\,)^2} \nonumber\\
 &= \frac{1}{\sum_{l=1}^{N}\frac{\rho^{-\alpha_l}}{(\rho^{-\alpha_l}+c\,)^2}} \frac{ \rho}{N}\sum_{l=1}^{N} \frac{\rho^{-2\;\alpha_l}}{(\rho^{-\alpha_l}+c\,)^2}\nonumber\\
 &\doteq \frac{1}{ \rho^{-\alpha_{\min}}}  \frac{\rho}{N} \rho^{-2\;\alpha_{\min}}\nonumber\\
&= \frac{1}{N} \rho^{1-\alpha_{\min}} . \label{eq:PTSimplified}
\end{align}

Using the union bound~\eqref{eq:PEPAsym},
\begin{align}
P(E,\bar{O})&\;\dot{\leqslant}\; 
 \begin{cases}
2^{Rl}e^{-\rho^{\alpha_{\min}}} &\text{$\alpha_{\min} < 1$ } \\
2^{Rl}e^{-\frac{\rho}{N}} &\text{$\alpha_{\min} > 1$ }
\end{cases} \label{eq:RLAlphamin}
\end{align}

Since the exponential function dominates polynomials we have
\begin{equation}
\lim_{\rho\to\infty} \frac{e^{-\rho^{\alpha_{min}}}}{\rho^{-MN}} = 0
\nonumber
\end{equation}
and
\begin{equation}
\lim_{\rho\to\infty} \frac{e^{-\rho}}{\rho^{-MN}} = 0
\nonumber
\end{equation}
which in turns gives
\begin{align}
P(E,\bar{O})\;\dot{\leqslant}\;  \rho^{-MN} \; . \label{eq:ProbEO}
\end{align}

Using~\eqref{eq:UpperBoundSum3} and~\eqref{eq:ProbEO}, the PEP
given by~\eqref{eq:PEPUB} is bounded as
\begin{align}
P_{\text{err}}&\;\dot{\leqslant}\; P_{out} +\;  P(E,\bar{O}) \nonumber\\
 &\;\dot{\leqslant}\; P_{out}^b +\; \rho^{-MN}\nonumber\\
&\doteq  P_{out}^b \nonumber\\
&= \rho^{-d_{out}}. \label{eq:UBOnPEP}
\end{align}
therefore $d\geqslant d_{out}$ which concludes the proof for the RZF system.

For the ZF precoding system, it can be directly shown that a similar
proof holds for both ZF precoding designs.

\subsection{Proof of Eq.~\protect\eqref{eq:ProbRZFDMT2}
}
\label{appendix:A}

Recall that
\[
\psi \triangleq \frac{1}{| u_{1l'} u^*_{2l'}|^2}+ \sum_{k=2}^{N} \frac{1}{| u_{kl'} u^*_{1l'}|^2}.
\]
All terms of $\psi$ the common factor
$\frac{1}{|u_{1l'}|^2}$. Thus we have
\begin{align}
\psi&=\psi_a\psi_b  \nonumber\\
\psi_a&=\frac{1}{|u_{1l'}|^2}\nonumber\\
 \psi_b&= \bigg( \frac{1}{|u^*_{2l'}|^2}
+\frac{1}{|u_{2l'}|^2}+ \frac{1}{|u_{3l'}|^2}+
\frac{1}{|u_{4l'}|^2}+\dots + \frac{1}{|u_{Nl'}|^2} \bigg). \label{eq:PsiElements}
\end{align}
Observe that all the terms of $\psi_b$ are distinct except for the
first two.

We now bound the probability $\prob \big(\psi \leqslant
\rho^{\frac{r}{N}} \big)$. 
\begin{align}
\prob \big(\psi \leqslant \rho^{\frac{r}{N}} \big) &\geqslant \prob \big(\psi \leqslant \rho^{\frac{r}{N}}\;|\;\psi<c\big)\; \prob(\psi<c)\nonumber\\
&\geqslant \prob \big(c \leqslant \rho^{\frac{r}{N}}\big)\; \prob(\psi<c)\nonumber\\
&\doteq \quad\prob(\psi<c)\label{eq:AsymptoticPsi}
\end{align}
Using $\psi=\psi_a\psi_b$ we can further
bound~\eqref{eq:AsymptoticPsi}
\begin{align}
\prob(\psi<c) &= \prob(\psi_a \psi_b<c)\nonumber\\
 &\geqslant \prob \big(\psi_a\psi_b \leqslant c\;\big|\psi_a<c_2\big)\; \prob(\psi_a<c_2)\nonumber\\
&\geqslant \prob \big(c_2\psi_b \leqslant c \big)\; \prob(\psi_a<c_2)\nonumber.
\end{align}
We thus have
\begin{align}
\prob \big(\psi \leqslant \rho^{\frac{r}{N}} \big) \geqslant \prob \big(\psi_b \leqslant c' \big)\; \prob(\psi_a<c_2)\label{eq:AsymptoConstant}
\end{align}
and $c'=c/c_2$.

We now evaluate the two probabilities in the right hand side of
~\eqref{eq:AsymptoConstant}. The first probability $\prob\big(\psi_b
\leqslant c' \big) = O(1) $. The proof easily follows
from~\cite[Appendix A]{Kumar:JIT09} with the observation that this
proof holds even when the two first elements of $\psi_b$ are the same.
The second probability $\prob(\psi_a<c_2)$ is evaluated as follows.
Let $q=|u_{1l'}|^2$. We use the following distributions
from~\cite[Appendix A]{Hochwald:TCOMM05}
\begin{align}
f(q)&=(N-1)(1-q)^{N-2}, \quad\, 0 \leqslant q \leqslant 1\nonumber
\end{align}
then
\begin{align}
\prob(\psi_b <c_2) &= \prob(q>\frac{1}{c_2})\nonumber\\
 &= \int_{\frac{1}{c_2}}^{1} f(q) \; dq \nonumber\\
&=(1-\frac{1}{c_2})^{N-2}   \label{eq:Integral1}
\end{align}

Observing that~\eqref{eq:Integral1} is not a function of $\rho$
concludes the proof.


\subsection{Proof of $\prob \big(\lambda_{l} \geqslant \xi \big) = O(1)$  for any $l$}
\label{appendix:ExponentialInequality}

Define a Wishart matrix $\mathbf W$ using the Gaussian matrix $\bH$.
\[
{\mathbf W}=
 \begin{cases}
 \bH\,\bH^H &\text{$M > N$ } \\
 \bH^H\bH &\text{$N \leqslant N$ }
\end{cases}.
\]

Let $n=\max(M,N)$ and $m=\min(M,N)$.  The matrix $\mathbf W$ is $m\times m$ random non-negative definite that has real, non-negative eigenvalues with $\lambda_1 \geqslant \dots \geqslant \lambda_m0$. The joint density of the ordered eigenvalues is~\cite{Telatar:EUR99}
\begin{equation}
f({\boldsymbol \lambda}) = K_{m,n}^{-1} e^{-\sum_i \lambda_i} \prod_i \lambda^{n-m}_i \prod_{i<j} (\lambda_i - \lambda_j)^2. \label{eq:OrderedEigenDist}
\end{equation}

Thus the marginal distribution of $\lambda_{l}$ is given
by~\cite{Telatar:EUR99}
\begin{align}
f_{{\boldsymbol \lambda}_l}(\lambda_{l}) &= \int \dots \int f({\boldsymbol \lambda}) \; d\lambda_2\dots d\lambda_m \nonumber\\
&= \frac{1}{m} \sum^{m}_{i=1} \varphi_i(\lambda_l)^2\lambda_l^{n-m} e^{-\lambda_1} \nonumber
\end{align}
where
\begin{equation}
\varphi_{k+1}(\lambda) =\bigg[ \frac{k!}{(k+n-m)!}  \bigg]^{1/2} L_k^{n-m}(\lambda), \;\; k=0,\dots,m-1 \nonumber
\end{equation}
where $L_k^{n-m}(x)=\frac{1}{k!} e^{x} x^{m-n} \frac{d^k}{dx^k} (e^{-x} x^{n-m+k})$ (with $L_0=1$) is the associated Laguerre polynomial of order $k$. 

We now compute $\prob \big(\lambda_{l} \geqslant \xi \big)$,
\begin{align}
\prob \big(\lambda_{l} \geqslant \xi \big) &= \int_{\xi}^{\infty}  \frac{1}{m} \sum^{m}_{i=1} \varphi_i(\lambda_l)^2\lambda_1^{n-m} e^{-\lambda_l} d\lambda_l \nonumber\\
&\geqslant \int_{\xi}^{\infty}  \frac{1}{m}  \varphi_1(\lambda_l)^2\lambda_l^{n-m} e^{-\lambda_l} d\lambda_l \nonumber\\
&= \int_{\xi}^{\infty}  \frac{1}{m(n-m)!}  \lambda_l^{n-m} e^{-\lambda_l} d\lambda_l \nonumber\\
&= \frac{1}{m(n-m)!}  \bigg( -e^{-\lambda_l} \lambda^{n-m}_l \; - \nonumber\\
&\hspace{20pt} e^{-\lambda_l} \sum_{k=1}^{n-m} n(n-1)\dots(n-k+1) \lambda^{n-m-k}  \bigg) \bigg|^{\infty}_{\xi} \label{eq:Integration}\\
&= \frac{1}{m(n-m)!}  \big( e^{-\xi} \xi^{n-m} \; + \nonumber\\
&\hspace{20pt} e^{-\xi} \sum_{k=1}^{n-m} n(n-1)\dots(n-k+1) \xi^{n-m-k}  \big)  \label{eq:Integration2}
\end{align}
where~\eqref{eq:Integration} follows from~\cite[Section 2.32]{Gradshtey:book}. The right hand side of Equation~\eqref{eq:Integration2} is a non-zero constant bounded away from zero. This concludes the proof.

\bibliographystyle{IEEEtran} \bibliography{IEEEabrv,AHesham}

 \end{document}